\documentclass[12pt,preprint]{aastex}

\newcommand{\te}{\emph{$t^2$}}

\shorttitle{ The \ion{H}{2} region NGC 3576}
\shortauthors{Garc\'{\i}a-Rojas et al.}

\received{}
\begin{document}

\title{Chemical abundances of the Galactic \ion{H}{2} region NGC 3576 derived from 
VLT echelle spectrophotometry\footnotemark{}}

\author{Jorge Garc\'{\i}a-Rojas} 
\affil{Instituto de Astrof{\'\i}sica de Canarias, E-38200, La Laguna, Tenerife, Spain}
\email{jogarcia@ll.iac.es}

\author{C\'esar Esteban}
\affil{Instituto de Astrof{\'\i}sica de Canarias, E-38200, La Laguna, Tenerife, Spain}
\email{cel@ll.iac.es}

\author{Manuel Peimbert}
\affil{Instituto de Astronom\'\i a, UNAM, 
Apdo. Postal 70-264, M\'exico 04510 D.F., Mexico}
\email{peimbert@astroscu.unam.mx}

\author{M\'onica Rodr\'{\i}guez}
\affil{Instituto Nacional de Astrof\'{\i}sica, \'Optica y Electr\'onica, Apdo. Postal 51 y 216, 72000 Puebla, Mexico}
\email{mrodri@inaoep.mx}

\author{Mar\'{\i}a Teresa Ruiz}
\affil{Departamento de Astronom\'{\i}a, Universidad de Chile,
Casilla Postal 36D, Santiago de Chile, Chile}
\email{mtruiz@das.uchile.cl}

\and

\author{Antonio Peimbert}
\affil{Instituto de Astronom\'\i a, UNAM, 
Apdo. Postal 70-264, M\'exico 04510 D.F., Mexico}
\email{antonio@astroscu.unam.mx}

\begin{abstract}
We present echelle spectrophotometry of the Galactic \ion{H}{2} region NGC 3576. The data have 
been taken 
with the VLT UVES echelle spectrograph in the 3100 to 10400 \AA\ range.
We have measured the intensities of 458 emission lines, 344 are permitted lines of H$^0$, 
He$^0$, C$^{+}$, N$^{0}$, N$^{+}$, N$^{++}$, O$^{0}$, O$^{+}$, Ne$^{+}$, S$^{++}$, Si$^{0}$, 
Si$^{+}$, Ar$^{0}$ and Ar$^{+}$; some of them are produced by recombination and others mainly by 
fluorescence. Electron 
temperatures and densities have been determined using different continuum and line intensity 
ratios.
We have derived He$^{+}$, C$^{++}$, O$^{+}$, O$^{++}$ and Ne$^{++}$ ionic abundances from pure 
recombination lines. 
We have also derived 
abundances from collisionally excited lines for 
a large number of ions of different elements. Remarkably consistent 
estimations of \emph{ $t^2$} have been obtained by comparing Balmer and Paschen to 
[\ion{O}{3}] temperatures, and O$^{++}$ and Ne$^{++}$ ionic abundances obtained from 
collisionally excited and recombination lines.
The chemical composition of NGC 3576 is compared with those of other Galactic 
\ion{H}{2} regions and with the one from the Sun. 
A first approach to the gas-phase Galactic radial abundance gradient of C as well as of the C/O ratio
has been made.
\end{abstract}


\footnotetext[1]{Based on observations collected at the European Southern
Observatory, Chile, proposal number ESO 68.C-0149(A).}

\section{Introduction
\label{intro}}

NGC 3576 ---also known as Gum38a--- comprises the western part of the RCW 57 complex \citep{rodg60} and 
corresponds to a bright knot embedded 
in a large system of diffuse emission gas filaments \citep{girardi97}. This knot is one of the 
most luminous Galactic \ion{H}{2} regions 
in the infrared \citep{figue02}. It was thought that most of the ionization
of NGC 3576 is due to two O stars (HD 97319 and HD 97484) 
and two B stars (HD 974999 and CPD--60$^{\circ}$2641) which are the main visual components of the OB 
association \citep{hum78}; 
but recent infrared observations suggest that the main ionizing sources of this \ion{H}{2} region 
are behind heavily obscuring clouds 
\citep{bor97, figue02}. It is located in Carina at a distance of 2.7 kpc \citep{russ03} and at a
Galactocentric distance of 7.4 kpc (assuming a Galactocentric solar distance of 8.0 kpc). 

Previous abundance determinations for NGC 3576 are those by \citet{girardi97}, based on the 
analysis of collisionally excited lines (hereafter CELs);
\citet{tsa03} based on CELs and some recombination lines (hereafter RLs) of C$^{++}$ and O$^{++}$; 
and \citet{simpson95} based on far-infrared data and 
photoionization models.

The temperature fluctuations problem \citep{peim67} is, nowadays, a  much-discussed topic in 
astrophysics of gaseous nebulae
\citep{liu02, liu03, este02, tor03}.
Traditionally, the abundance studies for \ion{H}{2} regions have been based on determinations from 
CELs, 
which are strongly dependent on the temperature 
variations over the observed volume of the nebula. Alternatively, RLs are almost independent of 
such variations and are, 
in principle, more precise 
indicators of the true chemical abundances of the nebula. Several authors have obtained 
O$^{++}$/H$^+$ from \ion{O}{2} recombination line intensities for the brightest \ion{H}{2} 
regions of the Galaxy (Peimbert, Storey \& Torres-Peimbert 
1993; Esteban et al. 1998, hereafter
EPTE; Esteban et al. 1999a, hereafter EPTGR; Esteban et al. 1999b; Tsamis et al. 2003), for 
extragalactic \ion{H}{2} regions \citep{este02b, peim03, tsa03} and for planetary
nebulae \citep{liu00, liu01, 
ruiz03, peim04}, and all of them have found that the abundance determinations from RLs are 
systematically larger than those obtained using CELs. 
The CEL abundances depend strongly on the adopted temperature while the RL abundances are 
almost independent of it. In the presence of temperature
inhomogeneities the temperature derived from the [\ion{O}{3}] diagnostic lines,
$T_e$(\ion{O}{3}), is considerably higher than the average one and than 
those temperatures derived from the Balmer and Pashen continua. 
For \ion{H}{2} regions the differences in both the abundances derived from CELs and RLs 
and the temperatures derived from CELs and recombination processes,
can be consistently accounted for by assuming a \te\ (mean square
temperature variation over the observed volume) in the range 0.020--0.044. 

\citet[][ see also Rubin et al. 2003]{odell03} have used a different 
method to show that there are temperature inhomogeneities in \ion{H}{2} regions: 
these authors have determined the columnar temperature along $1.5 \times 10^6$ 
lines of sight in the Orion nebula, the distribution of temperatures of 
their sample supports the $t^2$ values derived from other methods.
The origin of temperature fluctuations is still controversial and a serious challenge to our 
knowledge of the physics and structure of ionized nebulae.

We have taken long-exposure high-spectral-resolution spectra with the VLT UVES echelle 
spectrograph to obtain accurate measurements of very 
faint permitted lines of heavy element ions in NGC 3576. We have determined 
the physical 
conditions and the chemical abundances of NGC 3576 with high accuracy, including important 
improvements over previous determinations.
We have considered C$^{++}$ and O$^{++}$ abundances obtained from several permitted lines  of 
\ion{C}{2}
and \ion{O}{2}, avoiding the problems of line blending, including several 3d-4f transitions which 
are very useful for abundance 
determinations because they are free of optical depth effects (Liu et al. 1995 and references 
therein). 
We have also derived O$^+$ and Ne$^{++}$ abundances from RLs for the first time in this
nebula. We have computed 
$t^2$ values from the determination of the Balmer and Paschen temperatures,  
which coincide with the \te\ that produces the agreement 
between the ionic abundances 
obtained from CELs and RLs. Finally, we have determined helium abundances taking into account 
a large number of singlet lines of \ion{He}{1}.

In \S\S~\ref{obsred} and~\ref{lin} we describe the observations and the data reduction procedure. 
In \S~\ref{phiscond} we obtain temperatures and densities 
using several diagnostic ratios; also, in this section, we determine \te\ from different line 
intensity ratios and
temperature determinations. In \S~\ref{helioabund} we briefly analyze the recombination spectra of  
\ion{He}{1} and derive the He$^{+}$/H$^{+}$ ratio. In \S~\ref{ionic} ionic abundances are 
determined based on RLs, as well as on
CELs. In \S~\ref{abuntot} the total abundances are determined. In \S\S~\ref{discus} 
and ~\ref{conclu} we present the discussion and the conclusions, respectively.

\section{Observations and Data Reduction
\label{obsred}}

The observations were made on 2002 March 11 with the Ultraviolet Visual Echelle Spectrograph, UVES 
\citep{dodo00}, 
at the VLT Kueyen Telescope in Cerro Paranal Observatory (Chile). We used the standard settings in 
both the red and 
blue arms of the spectrograph, covering the region from 3100 to 10400 \AA\ . The log of the 
observations is presented in
Table~\ref{tobs}.

The wavelength regions 5783--5830 \AA\ and 8540--8650
\AA\ were not observed due to a gap between the two CCDs used in
the red arm. There are also five small gaps that were not observed, 9608--9612 \AA, 
9761--9767 \AA, 9918--9927 \AA, 
10080--10093 \AA\ and 10249--10264 \AA, because  
the five redmost orders did not fit completely within the CCD.  We took long and short exposure 
spectra to check for possible saturation effects.

The slit was oriented east-west and the
atmospheric dispersion corrector (ADC) was used to keep the same observed
region within the slit regardless of the air mass value.  The slit width was
set to 3.0$\arcsec$ and the slit length was set to 10$\arcsec$ in the blue arm and to 12$\arcsec$ in
the red arm; the slit width was chosen to maximize the S/N ratio of the
emission lines and to maintain the required resolution to separate most of the
weak lines needed for this project. The effective resolution for the NGC 3576 lines
at a given wavelength is approximately $\Delta \lambda \sim \lambda / 8800$. 
The center of the slit was placed 65$\arcsec$ west and 24 $\arcsec$ north 
of HD 97499, covering the brightest region of NGC 3576. The reductions were made for an area 
of 3$\arcsec$$\times$8.25$\arcsec$ in the blue arm (B1 and B2), 3$\arcsec$$\times$10.1$\arcsec$ in the red arm (R1) and 
3$\arcsec$$\times$9.5$\arcsec$ in the R2 configuration of the red arm. These areas were chosen in 
order to maximize the S/N ratio of the emission lines. A test was made to confirm that 
line fluxes were not significantly affected by the different areas chosen in each 
spectral range. 
 
The spectra were reduced using the IRAF\footnotemark{} echelle reduction
package, following the standard procedure of bias subtraction, aperture
extraction, flatfielding, wavelength calibration and flux calibration. The standard star EG~247 
was observed for flux
calibration.

\footnotetext{IRAF is distributed by NOAO, which is operated by AURA,
under cooperative agreement with NSF.}

\section{Line Intensities and Reddening Correction
\label{lin}}

Line intensities were measured integrating all the flux in the line between two 
given limits and over a local continuum estimated by eye. In the cases of line blending, 
a multiple Voigt profile fit procedure was applied to obtain the line flux of each 
individual line. Most of these measurements were made with the SPLOT routine of the IRAF 
package. In some cases of very tight blends or blends with very bright telluric lines the 
analysis was performed via Gaussian fitting making use of the Starlink DIPSO software 
\citep{how90}.

Table~\ref{lineid} presents the emission line intensities of NGC 3576. The
first and fourth columns include the adopted laboratory wavelength, $\lambda_0$, and 
the observed wavelength in the heliocentric framework, $\lambda$. 
The second and third columns include the ion and the multiplet number, or 
series for each line.  The fifth and sixth columns include the observed
flux relative to H$\beta$, $F(\lambda$), and the flux corrected for reddening
relative to H$\beta$, $I(\lambda$). The seventh column includes the
fractional error (1$\sigma$) in the line intensities (see \S~\ref{errors}).

A total of 458 emission lines were measured; of them 344 are permitted, 108 are forbidden and 2 
are semiforbidden (see Table~\ref{lineid}). Some [\ion{N}{1}] and [\ion{O}{1}] lines were 
detected, but they are blended with telluric lines, making impossible their measurement. Several 
other lines were strongly affected by atmospheric features in absorption or by internal 
reflections by charge transfer in the CCD, rendering their intensities unreliable. Also, 29 lines 
are dubious identifications and 4 emission lines could not be identified in any of the available references.
Those lines are indicated in Table ~\ref{lineid}.

The identification and adopted laboratory wavelengths of the lines were obtained following 
previous identifications in 
the Orion Nebula by EPTE, \citet{bald00} and \citet{ost92}; we also 
used the compilations of atomic data by \citet{moo45,moo93}, \citet{wie66}, the 
line catalogue for gaseous nebulae of \citet{fek94}, the catalogue of \citet{peq88} for 
\ion{He}{1} lines and the papers of EPTGR on M8, \citet{EPTG} on M17 and \citet{liu00, liu01} 
on spectrophotometry of the planetary nebulae NGC 6153, M 1-42 and M 2-36. Also, we have used an
interactive source of nebular data: The Atomic Line List 
v2.04\footnote{webpage at: http://www.pa.uky.edu/$\sim$peter/atomic/}, directly or through the {\emph{
EMILI}\footnote{webpage at: http://www.pa.msu.edu/astro/software/emili/}} code \citep{shar03}

Following \citet{girardi97} we have assumed the extinction law of \citet{sav79} with R$_v$=3.1. 
A reddening coefficient of $C(H\beta)$=1.40$\pm$0.07 dex was determined by fitting the observed 
$I$(H Balmer lines)/$I$(H$\beta$) ratios (from H16 to H$\beta$) and $I$(H Paschen 
lines)/$I$(H$\beta$) 
(from P22 to P7), to the theoretical 
ones computed by Storey \& Hummer (1995) for $T_e$ = 9000 K and $N_e$ = 1000 cm$^{-3}$ (see 
below). \ion{H}{1} lines affected 
by blends or atmospheric absorption were not considered.

\subsection{Errors
\label{errors}}

The observational errors associated with the line intensities have been determined taking 
into account 
two sources of uncertainty: statistical errors in the line flux measurements and $C$(H$\beta$) 
uncertainty. It has not been possible to determine the systematic error of the flux 
calibration because we used a single standard star (EG 274). However, the comparison between 
observed and theorical Balmer and Paschen line ratios of the brightest --and no sky-affected-- lines show that 
the average differences are below 3\%. Moreover, in a future paper (Esteban et al., in preparation), we compare the  
echelle observations of EPTE with new VLT data for the same zone of the Orion nebula. These VLT data have been 
flux calibrated in identical form than our observations of NGC~3576 and do not show any systematical 
differences in the emission line ratios in common with EPTE, which differ typically not more than 3\%.  
Therefore, we can conclude that the flux calibration of the data presented in the present paper is 
confident, and it is not a source of significant systematical uncertainties.

The method developed to 
determine the line uncertainties consist of the following steps: firstly, the spectral ranges with 
the same exposure time (B1 and R1; B2 and R2) are grouped, then in each of these 
ranges several lines covering all the intensity ranges are chosen and the statistical errors 
are computed using the IRAF SPLOT task.
Error propagation theory and a logarithmic interpolation of $F$($\lambda$)/$F$(H$\beta$) \emph{ vs.} 
$\sigma$($F$($\lambda$)/$F$(H$\beta$)) 
are used to determine $\sigma$($F$($\lambda$)/$F$(H$\beta$)) for each line. Taking into account the 
uncertainties in the determination of
$C$(H$\beta$) and error propagation, the final percentile errors (1$\sigma$) of the 
$I(\lambda$)/$I({\rm H}\beta)$ ratios are computed and presented in column 7 of 
Table~\ref{lineid}. Colons indicate errors higher than 40 \%.

\section{Physical Conditions
\label{phiscond}}

\subsection{Temperatures and Densities
\label{temden}}

The large number of emission lines identified and measured in the spectra allows us the derivation 
of physical conditions using different 
line ratios. The temperatures and densities are presented in Table~\ref{plasma}. Most of the 
determinations were carried out with the 
IRAF task TEMDEN of the package NEBULAR, based in the FIVEL program developed by \citet{derob87} 
and improved by \citet{shaw95}. 

A representative initial $T_e$ of 10000 K  
was assumed in order to derive $N_e$(\ion{O}{2}), $N_e$(\ion{S}{2}), $N_e$(\ion{Cl}{3}) and 
$N_e$(\ion{Ar}{4}). On the other hand, 
we have derived the [\ion{Fe}{3}] density from the intensity of 14 lines, that seem 
not to be affected by line blending, together with the 
computations of \citet{rodri02}. The computed value is very consistent with the densities derived 
from [\ion{Cl}{3}] and [\ion{Ar}{4}] 
lines. From Table~\ref{plasma} it can be seen that all diagnostic ratios are in good agreement, 
except the densities obtained from [\ion{O}{2}] and [\ion{S}{2}], which give lower values than 
the other indicators. 

We have derived $N_e$(\ion{O}{2}) using the classical ratio
$\lambda$3726/$\lambda$3729. In spite of the low critical density of the highest level of 
the $\lambda$3729 transition, this ratio does not saturate in NGC 3576, but with this density we 
obtained a value of $T_e$(\ion{O}{2})=10800 K which is too high if compared with  
$T_e$(\ion{S}{2}) and $T_e$(\ion{N}{2}). Due to the extreme sensitivity of $T_e$(\ion{O}{2})
with the adopted density, we have decided to use the
$\lambda\lambda$3726+3729/$\lambda\lambda$7320+7330 ratio to derive
a new  $N_e$(\ion{O}{2}), because the abundances obtained with the different individual lines of
[\ion{O}{2}] assuming this density and $T_e$(\ion{N}{2}) are more consistent than those
obtained with the density derived from the usual ratio. It is not the aim of this work to solve 
this problem but it could be due to several reasons including errors in the \ion{O}{2} atomic data 
and the uncertainty in the contribution of recombination to the excitation of the auroral lines.  
To obtain  $N_e$(\ion{O}{2}) it is necessary to subtract the contribution to 
$\lambda\lambda$7320+7330 due to recombinations;
 \citet{liu00} find that the contribution to the 
intensities of the [\ion{O}{2}] $\lambda\lambda$ 
7319, 7320, 7331, and 7332 lines due to recombination can be fitted in the range 
0.5$\le$T/10$^4$$\le$1.0 by:
\begin{equation}
\frac{I_R(7320+7330)}{I({\rm H\beta})}
= 9.36\times(T_4)^{0.44} \times \frac{{\rm{O}}^{++}}{{\rm{H}}^+},
\end{equation}
where $T_4$=$T$/10$^4$. With this equation we estimate a contribution of 
approximately 6\% to the observed line intensities.

A weighted average of $N_e$(\ion{O}{2}), $N_e$(\ion{Fe}{3}), $N_e$(\ion{Cl}{3}) and 
$N_e$(\ion{Ar}{4}) was then used to derive $T_e$(\ion{N}{2}), 
$T_e$(\ion{O}{3}), $T_e$(\ion{Ar}{3}) and $T_e$(\ion{S}{3}), and iterated until convergence. So, 
for all the species except for S$^+$ the adopted value for the density is: $N_e$=2800$\pm$400 
cm$^{-3}$.

For S$^+$ we have adopted $N_e$(\ion{S}{2})=1300$^{+500}_{-300}$ cm$^{-3}$, because this ion has 
the lowest ionization potential of all the species studied. The $T_e$(\ion{S}{2}) derived making 
use of this density is much more consistent with temperatures derived using other diagnostic 
ratios than that derived with $N_e$=2800 cm$^{-3}$, which gives a temperature 
2000 K lower.

\citet{liu00} determined that the contribution to the intensity of the $\lambda$ 5755 
[\ion{N}{2}] line due
to recombination can be estimated from:
\begin{equation}
\frac{I_R(5755)}{I({\rm H\beta})}
= 3.19\times(T_4)^{0.30} \times \frac{{\rm N}^{++}}{{\rm H}^+},
\end{equation}
in the range 0.5$\le$ $T$/10$^4$$\le$2.0. We have obtained a contribution of recombination of about 
7.5\%, that represents a
decrease of more than 200 K in the temperature 
determination. 

Finally, considering the similarity of the temperature determinations based on CELs, an 
average of [\ion{O}{3}] and [\ion{N}{2}] temperatures was adopted for NGC 3576, 
assuming a 1-zone ionization scheme. This is because these diagnostic temperatures are the usually 
adopted ones to characterize high and low ionization zones respectively, and because in our case 
they are coincident. We obtain a representative value of $T_e$=8500$\pm$150 K. $T_e$(\ion{S}{2}) and 
$T_e$(\ion{Ar}{3}) temperatures are absolutely consistent with the adopted temperature and do not affect 
to the average.
We have not included $T_e$(\ion{S}{3}) because this is the most discrepant value.

The Balmer continuum temperature was determined following the equation by \citet{liu01}:
\begin{equation}
T({\rm Bac}) = 368 \times(1 + 0.259y^+ + 3.409y^{++})\left( \frac{{\rm Bac}}{{\rm 
H11}}\right)^{-3/2} ~{\rm {K}};
\end{equation}
where $y^+$ and $y^{++}$ are the He$^+$/H$^+$ and He$^{++}$/H$^+$ ratios respectively, 
and $Bac$ is the value of the discontinuity 
of the balmer jump in erg cm$^{-2}$ s$^{-1}$ \AA$^{-1}$.
A power-law fit to the relation between $I_c$($Bac$)/$I$(H$n$) and $T_e$ for 
$3\le n\le 20$ gives the same result as 
the method mentioned above, but higher uncertainties due to the statistical dispersion. 
The Paschen continuum temperature 
was derived fitting the relation between $I_c$($Pac$)/$I$(P$n$) and $T_e$ 
for $7\le n\le 25$. The emissivities as a function of the electron temperature for the nebular 
continuum and the \ion{H}{1} Balmer and Paschen lines were taken from \citet{brown70} and 
\citet{sto95} respectively. The finally adopted value of $T$($Pac$) was the average of 
those obtained using the different \ion{H}{1} lines, neglecting those which are affected 
by atmospheric features. 

\begin{figure}
\begin{center}
\rotate
\epsscale{1.4}
\plottwo{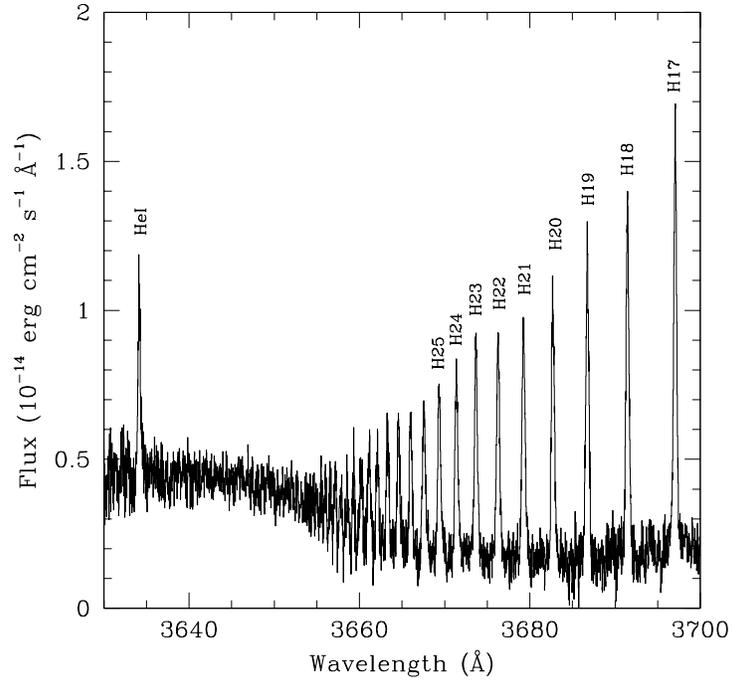}{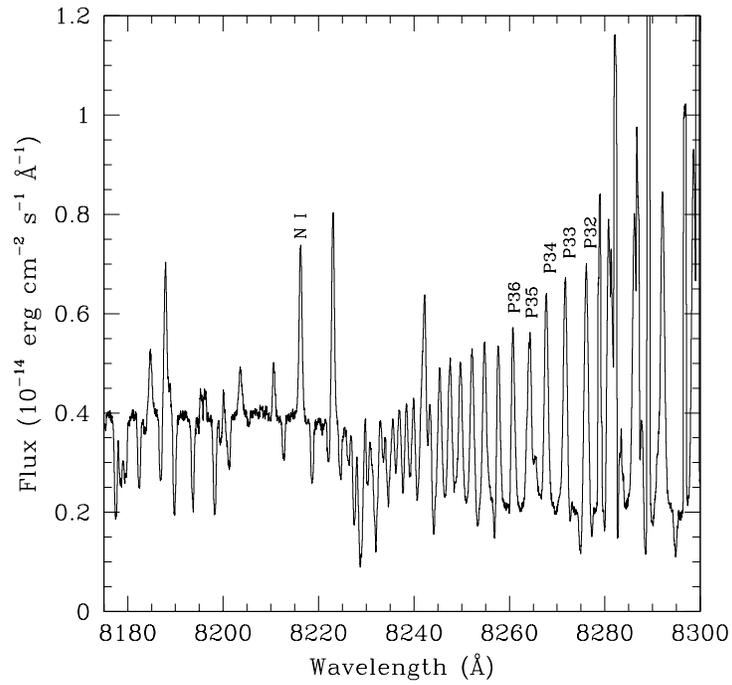}
\figcaption{Section of the echelle spectrum including the Balmer (top) and the Paschen (bottom) 
limits (observed fluxes).
\label{saltos}}
\end{center}
\end{figure}

Figure~\ref{saltos} shows the spectral regions near the Balmer and the Paschen limits. The 
discontinuities can be easily appreciated. 

\subsection{Temperature variations}

Under the assumption of a constant electron temperature, RLs of heavy elements 
yield higher abundance values relative to hydrogen than CELs. This is a well known result that 
different authors have corroborated for \ion{H}{2} regions and planetary nebulae (e.g. EPTE; 
Esteban et al. 2002; Liu et al. 2002, and references therein). \citet{peim67} 
proposed the presence of spatial temperature 
fluctuations (parameterized by $t^2$) as the cause of this 
discrepancy, because CELs and RLs intensities have different dependences on 
the electron temperature. In addition, and for the same reason, the 
comparison between $T$($Bac$) or $T$($Pac$) and 
electron temperatures obtained from forbidden line analysis can give an indication of such 
fluctuations. 

A complete formulation of temperature fluctuations has been developed by \citet{peim67}, \citet{peim69} 
and \citet{peim71}. 
To derive the value of $t^2$ we have followed \citet{peim00} and \citet{pepe02}. 
We have assumed a 1-zone ionization scheme, and have combined the temperature 
derived from the ratio of the [\ion{O}{3}] $\lambda\lambda$4363,
5007 lines with the temperature derived from the ratio of the Paschen continuum to $I$(H$\beta$),
$T$($Pac$), which are given by:
\begin{eqnarray}
T(}$\ion{O}{3}${)= T_{0} \left[ 1 + {\frac{1}{2}}\left({\frac{91300}{T_{0}}} - 3\right) t^2\right]
\label{to3},
\end{eqnarray}
and
\begin{equation}
T(Pac) = T_0 (1 - 1.67t^2)
\label{tpac};
\end{equation}
we have labeled the resulting value as $FL-Pac$ in Table~\ref{t2}. 
Similarly we have combined $T_e$(\ion{O}{3}) with the temperature derived from the ratio of the 
Balmer continuum to $I$(H$\beta$),
$T$($Bac$), which is given by:
\begin{equation}
T(Bac) = T_0 (1 - 1.67t^2)
\label{tbac};
\end{equation}
we have labeled the resulting value as $FL-Bac$ in
Table~\ref{t2}. 

On the other hand, we have derived the 
$t^2$ value that produces the agreement between the ionic abundances obtained from both 
recombination and forbidden lines for the O$^{++}$ ion. In particular, we
have found that the RL/CEL ratio for O$^{++}$ in this case is 1.9, which is in
excellent agreement with the value found by \citet{tsa03} that amounts to
1.8 (see \S~\ref{recom} for discussion about RLs abundances). Also, we have derived the $t^2$ 
value from the RL/CEL ratio of the Ne$^{++}$ abundance, which is completely 
consistent with the $t^2$ derived from the RL/CEL ratio of O$^{++}$ abundances. We have not considered 
the $t^2$ derived from the RL/CEL ratio of O$^{+}$ due to the high uncertainty of the abundance derived
from the only suitable RL in our spectra (see \S~\ref{recom})

The values of $t^2$ obtained are shown in Table~\ref{t2}.
We have adopted a final value of $t^2$=0.038$\pm$0.009, that is, the average of O$^{++}$(R/C), 
Ne$^{++}$(R/C), $T_e$($FL-Pac$), and $T_e$($FL-Bac$) $t^2$'s weighted by their uncertainties.
This value is consistent with the correlation
showed by \citet{liu00} ---their Figure 8--- between the ratio of CELs and RLs abundances with
$T_e$($FL-Bac$). This correlation supports the idea that the disparities between electron
temperatures and abundances, are closely related and probably have the same origin.

\section{He$^+$ abundance
\label{helioabund}}

There are 91 \ion{He}{1} emission lines identified in our spectra. These lines arise mainly from 
pure recombination, although they may have contributions from collisional excitation and self-absorption 
effects. On the other hand, singlets are, in general, more suitable for deriving an accurate the He$^{+}$/H$^{+}$
ratio, because they are not affected by self-absorption effects. Due to the large number of singlet lines
detected, and to their good signal-to-noise ratio, we have decided to derive the He$^+$/H$^+$ ratio 
making use of these lines.

To obtain He$^{+}$/H$^{+}$ values we need a set of effective recombination coefficients for the He 
and H lines, and to estimate the contribution due to collisional excitation to the helium line 
intensities (which is in fact rather small for singlet lines). The recombination coefficients used 
were those by \citet{sto95} for \ion{H}{1}, and those by \citet{smits96}
and \citet{ben99} for \ion{He}{1}. The collisional contributions were estimated from the 
computations by \citet{ben99}.

In the low-density and low optical depth limit the emissivities of the helium and
hydrogen lines are proportional to powers of the temperature and $T_e$(\ion{He}{2}) is given by
\citep{peim67}:
\begin{eqnarray}
\nonumber T(}$\ion{He}{2}${)& = & T_0[1+(\langle\alpha\rangle+\beta-1)t^2/2] \\
                               & = & T_0(1-1.3t^2),
\end{eqnarray}
where $\langle\alpha\rangle$ is the average value of the power of the temperature for the helium lines
that we have used to derive the He$^+$/H$^+$ ratio; $\langle\alpha\rangle$ it was
derived from \citet{ben99} and $\beta$ is the power of temperature for H$\beta$, 
derived from \citet{sto95}. With this relation we have derived a value of $T$(\ion{He}{2})=6800$\pm$400 K.

Table~\ref{abhe} presents the He$^{+}$/H$^{+}$ values obtained for the ten brightest and best 
observed helium singlet lines. We have obtained He$^{+}$/H$^{+}$ = 0.087$\pm$0.008 for
$T$(\ion{He}{2})=6800 K and $N_e$=2800 cm$^{-3}$.

\section{Ionic Abundances
\label{ionic}}

\subsection{Ionic Abundances from CELs
\label{cels}}

Ionic abundances of N$^+$, O$^+$, O$^{++}$, Ne$^{++}$, S$^+$, S$^{++}$, Cl$^+$, Cl$^{++}$, 
Cl$^{+3}$, Ar$^{++}$ and 
Ar$^{3+}$ have been determined from CELs, using the IRAF package NEBULAR (except for Cl$^+$, see 
below). 
Additionally, we have determined the ionic abundances of Fe$^{++}$ and Fe$^{3+}$ following the 
methods and data discussed below. As we have shown in \S~\ref{temden}, we have adopted an 
$N_e$($low$)=1300 cm$^{-3}$ for the ion with the lowest ionization potential: S$^+$,
and the same temperature for all the ions. Ionic abundances are listed in 
Table ~\ref{celabun} and correspond to the mean value of the 
abundances derived from all the individual lines of each ion observed (weighted by their relative 
strengths). The values obtained 
are very consistent with those derived by \citet{tsa03} for the ions in common (differences not 
larger than 0.15 dex). 

The Cl$^{+}$/H$^{+}$ ratio cannot be derived from the NEBULAR routines,
instead we have used an old version of the five-level atom program of
\citet{shaw95} 
that is described by \citet{derob87}. This version uses the atomic data for Cl$^{+}$ 
compiled by \citet{men83}. In any case, the atomic data for this ion and therefore 
the Cl$^{+}$/H$^{+}$ ratio derived are rather uncertain (Shaw 2003, personal 
communication).

To derive the abundances for $t^2$ = 0.038 we used the abundances for $t^2$=0.00 and the 
formulation of by \citet{peim67} and \citet{peim69} for $t^2>$0.00. 
To derive abundances for other $t^2$ values it is possible to interpolate or extrapolate 
the values presented in Table~\ref{celabun}.

Many [\ion{Fe}{2}] lines have been identified in the optical spectra of \ion{H}{2} 
regions (Rodr\'{\i}guez 1996; EPTE; EPTGR). Most 
of these lines are severely affected by fluorescence effects \citep{rodri99, ver00}. 
Unfortunately, we can not measure the [\ion{Fe}{2}] $\lambda$ 
8617 \AA\ line, which is almost insensitive to the effects of UV pumping. This line is precisely 
in one of the observational gaps of our spectroscopic configuration (see \S~\ref{obsred}). 
However, we do measure [\ion{Fe}{2}] $\lambda7155$, a line which is not much affected by fluorescence 
effects \citep{ver00}. We have derived an estimation of the Fe$^+$ abundance from this line 
assuming that $I(\lambda7155)/I(\lambda8616)\thicksim 1$ \citep{rodri96} and using the calculations of 
\citet{bau96}. We find Fe$^+$/H$^+$$\thicksim$3.5$\times$10$^{-8}$, a value much lower than the values obtained for 
the Fe$^{++}$ and Fe$^{3+}$ abundances (see below). Therefore, the Fe$^+$ abundance will be considered 
negligible in what follows.

The calculations for Fe$^{++}$ have been done with a 34 level model-atom that uses the 
collision strengths of \citet{zha96} and the transition probabilities of \citet{qui96}. 
We have used 14 [\ion{Fe}{3}] 
lines that do not seem to be affected by blends.
We find an average value and a standard deviation of 
Fe$^{++}$/H$^+$=(3.68$\pm$0.36)$\times$10$^{-7}$. Adding errors in $T_e$ and
$N_e$ we finally obtain 12+log(Fe$^{++}$/H$^+$)=5.57$\pm$0.05. 

We have detected [\ion{Fe}{4}] $\lambda6739.8$, the brightest optical [\ion{Fe}{4}] line for the 
physical conditions in NGC 3576. For deriving the Fe$^{3+}$/H$^{+}$ ratio. we have used a 33-level 
model-atom where all  collision strengths are those calculated by \citet{zha97}, the transition 
probabilities are those recommended by \citet{fro98} (and those from \citet{gar58} for the 
transitions not considered by \citet{fro98}).
Assuming an uncertainty of 50\% in the intensity measured, we have
derived a value of 12+log(Fe$^{3+}$/H$^+$)=5.71$^{+0.17}_{-0.29}$.

Values of Fe$^{++}$ and Fe$^{3+}$ abundances for t$^2>$0.00 are also shown in Table ~\ref{celabun}.

\subsection{Ionic Abundances from RLs
\label{recom}}

We have measured 170 permitted lines of heavy element ions such as \ion{O}{1}, \ion{O}{2}, 
\ion{C}{2}, 
\ion{Ne}{1}, \ion{Ne}{2}, \ion{S}{1}, \ion{S}{2}, \ion{N}{1}, \ion{N}{2}, \ion{N}{3}, 
\ion{Ar}{1}, \ion{Ar}{2}, \ion{Si}{1},and \ion{Si}{2}, most of them detected for the first time in 
NGC 3576.

As we noted in \S~\ref{intro}, those permitted lines produced by recombination can give accurate 
determinations of ionic abundances because their relative intensities depend weakly on electron 
temperature and density.

Let $I$($\lambda$) be the intensity of a recombination line of an element X, \emph{ i} times 
ionized at wavelength $\lambda$; then 
the abundance of the ionization state $i+1$ of element X is given by:
\begin{equation}
\frac{N({\rm X}^{i+1})}{N({\rm H}^+)}=\frac{\lambda(\rm\AA)}{4861}\frac{\alpha_{eff}({\rm 
H}\beta)}{\alpha_{eff}(\lambda)}
\frac{I(\lambda)}{I({\rm H}\beta)},
\end{equation}
where $\alpha_{eff}$($\lambda$) and $\alpha_{eff}$(H$\beta$), are the effective recombination 
coefficients for the line and
H$\beta$, respectively. The $\alpha_{eff}$($\lambda$)/$\alpha_{eff}$(H$\beta$) ratio is almost 
independent of the adopted temperatures and densities, and varies by less than a few percent 
within the temperature and density ranges presented in Table~\ref{plasma}.

Following EPTE we have taken into account the abundances obtained from the intensity of each 
individual line and the abundances
from the estimated total intensity of each multiplet, which is obtained by multiplying the sum of 
the intensities of the observed lines by the multiplet correction factor,
\begin{equation}
m_{cf}=\frac{\sum\limits_{{\rm all}\ i,j}s_{ij}}{\sum\limits_{{\rm obs}\  i,j}s_{ij}},
\end{equation}
where the upper sum runs over \emph{all} the lines of the multiplet, and the lower sum runs over 
the \emph{observed} lines of the multiplet. The theoretical line strengths, $s_{ij}$, are 
constructed assuming that they are proportional to the population of their parent levels assuming 
LTE computation predictions. The abundances derived by this manner are labeled as 
"Sum" in Tables~\ref{cii}, ~\ref{nitro}, ~\ref{oi}, ~\ref{oii} and ~\ref{neii}. This quantity corresponds to 
the expected abundance given by the whole multiplet. Abundances that we have taken into account 
are marked as boldface in Tables 7 to 11.

\begin{figure}
\begin{center}
\epsscale{1.0}
\plotone{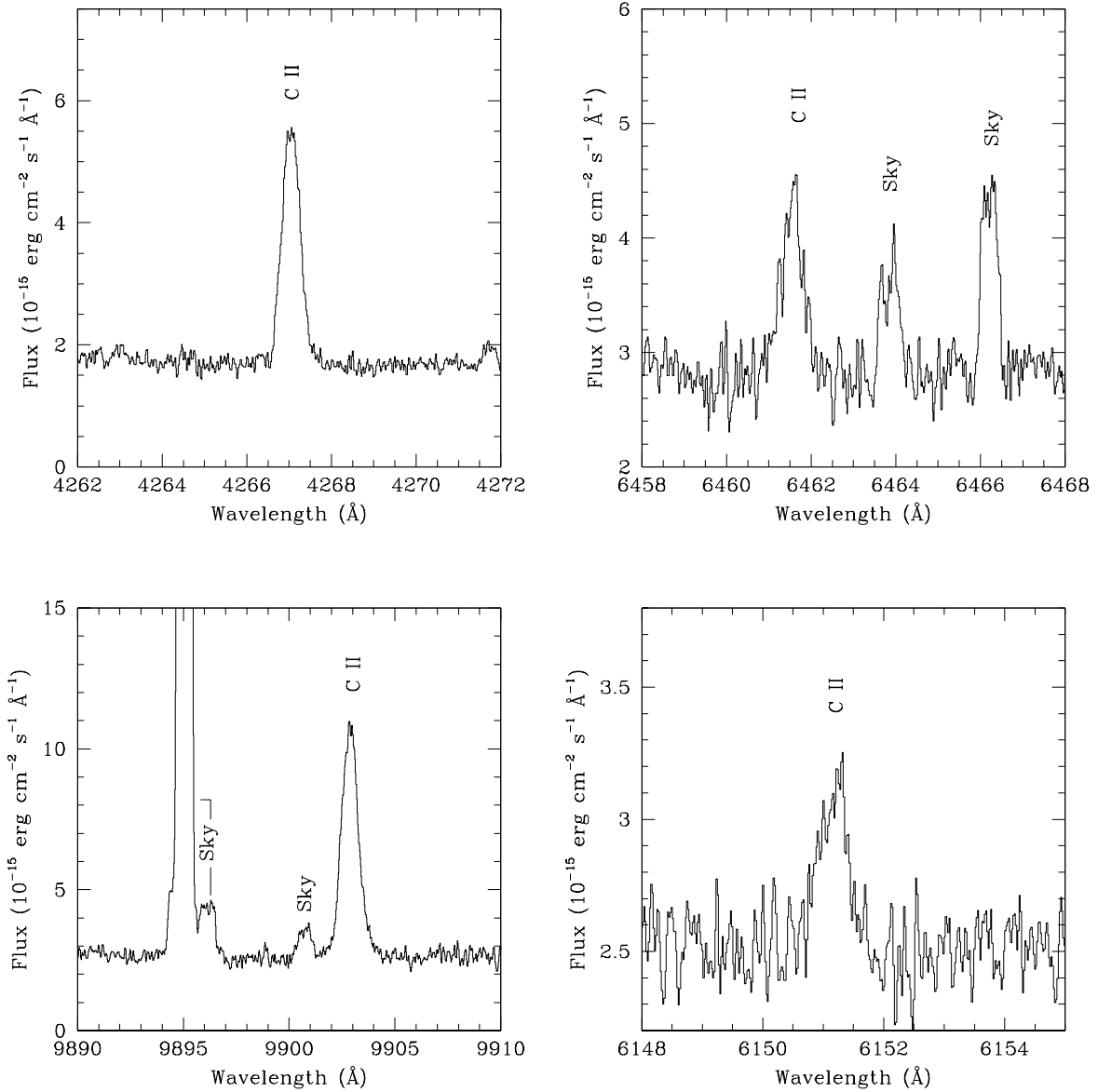}
\figcaption{Sections of the echelle spectrum of NGC 3576 showing the brighest lines of
\ion{C}{2} detected
\label{ciilines}}
\end{center}
\end{figure}

Ten permitted lines of \ion{C}{2} have been measured in NGC 3576. Some of these lines (those of 
multiplets 6, 16.04, 17.02, 17.04 and 17.06) 
are $3d-4f$ transitions and are, in principle, excited by pure recombination (see Grandi 1976). 
Also, the abundances obtained from them are 
case-independent, so we adopted the mean of the values obtained for these transitions as our 
final adopted C$^{++}$/H$^+$ ratio. In any case, the result for the case sensitive multiplet 3 
gives a C$^{++}$ abundance for case B consistent with that adopted. On the other hand, \ion{C}{2} 
$\lambda$6578.05 
is also case sensitive and considerably affected by a telluric line. In Table~\ref{cii} we 
summarize the abundances 
obtained from the different lines detected as well as the adopted average value. We used the 
effective recombination 
coefficients computed by \citet{dav00}.
The dispersion of the abundances obtained by the different lines is very small and the final 
result is in very good 
agreement with the value obtained by 
\citet{tsa03} from the \ion{C}{2} $\lambda$4267 line alone. Figure~\ref{ciilines} shows the
high signal--to--noise ratio of the four brighest \ion{C}{2} lines detected in our spectrum.

We have detected 13 lines of \ion{N}{1} of multiplets 1, 2 and 3. It is a well known result that 
starlight excitation is the main responsible
of the observed strength of these lines \citep{gra75a,gra75b}. The abundances obtained from \ion{N}{1} 
lines in NGC 3576 are between 150 to 400 times 
higher than the abundances obtained with CELs (see Table~\ref{nitro}), a clear indication that 
these lines are mainly produced
by starlight excitation and not by recombination.

Abundances obtained for N$^{++}$ are shown in Table~\ref{nitro}. \citet{gra76} has shown that 
resonance fluorescence by the recombination line \ion{He}{1} $\lambda$508.64 
is the dominant mechanism to excite the 4s$^3P^0$ term of \ion{N}{2} in the Orion Nebula, and 
hence it should be responsible for the strength of multiplets 3 and 5. 
The term 4f$^3F$ cannot be reached by permitted resonance transitions and, therefore, it is 
excited mainly by recombination, 
so the abundance obtained from the $\lambda$4239.4 line of multiplet 
48 has been considered. 
Also, \citet{gra76} suggests that multiplets 28 and 20 could be excited by a combination of 
starlight and recombination. The recombination coefficients used are from  
\citet{kiss02} for all multiplets except multiplet 48 for which we have adopted the recombination 
coefficients of 
\citet{esc90}. Multiplets 5, 20 and 28 are strongly case-sensitive, therefore we have adopted the 
value given by the average of multiplets 3 and 48 in case B as a representative value of 
N$^{++}$/H$^+$.

The O$^+$ abundance was derived only from the \ion{O}{1} $\lambda$7771.94 line, because the other 
lines of multiplet 1
were strongly affected by telluric lines, as well as the only line of multiplet 4 detected, 
\ion{O}{1} $\lambda$ 8446.48. 
Multiplet 1 is case-independent and is 
produced by recombination because it corresponds to a quintuplet transition (the ground level is a 
triplet). 
The effective recombination coefficients were obtained from two sources: \citet{peq91} and 
\citet{esc92}. 
Though the results are very similar, we adopted the mean of the abundances obtained with the two 
different coefficients. 
The O$^+$ abundance obtained from the $\lambda$7771.94 line is quite uncertain because it is partially 
blended with a sky
emission line.

\begin{figure}
\begin{center}
\epsscale{1.0}
\plotone{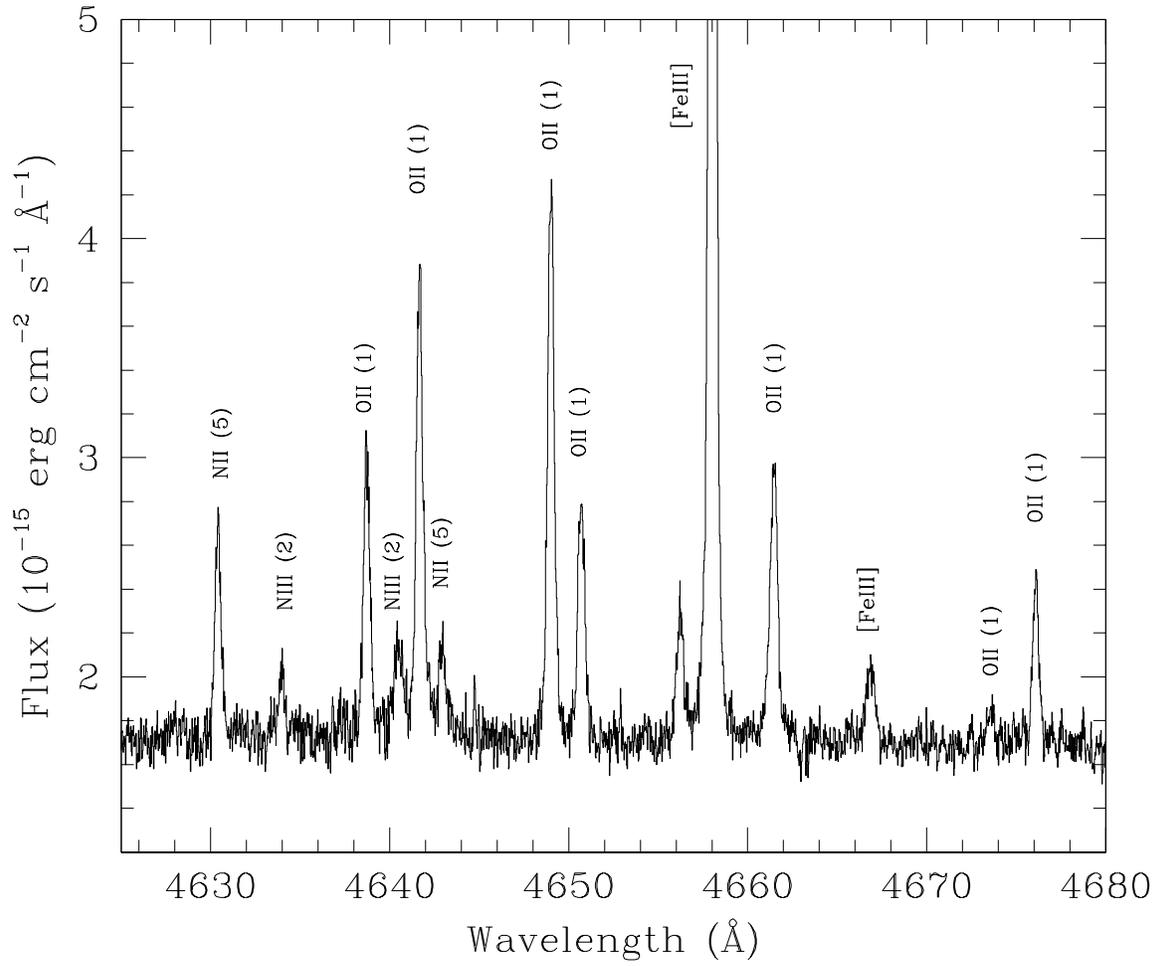}
\figcaption{Section of the echelle spectrum of NGC 3576 showing all the lines of multiplet 1 of 
\ion{O}{2}.
\label{m1oii}}
\end{center}
\end{figure}

More than 40 lines of \ion{O}{2} have been detected in our data. This is, along with that reported 
by \citet{este04} for Orion,  
one of the best \ion{O}{2} recombination-line spectrum that have been observed for a Galactic 
\ion{H}{2} region. O$^{++}$/H$^+$ ionic 
abundance ratios are presented in Table~\ref{oii}. Figure ~\ref{m1oii} shows the high quality of 
the spectrum in the zone of
multiplet 1 of \ion{O}{2}. This figure can be compared with Figure 3 of EPTE (Orion nebula), 
Figure 2 of \citet{peim03} (30
Doradus) and Figure 1 of \citet{este02b} (NGC 604). All these figures show the
same spectral zone and a direct comparison of
the quality of the spectra can be made.
Effective recombination coefficients are from \citet{sto94} for 3s-3p and 3p-3d
transitions 
--LS-coupling-- and from \citet{liu95} for 3p-3d and 3d-4f transitions
--intermediate coupling--, assuming 
case A for doublets and case B for quartets (for definitions of cases A, B and
C, see EPTE). For multiplet 15 we use the dielectronic 
recombination rate of \citet{nuss84}. The intensity of the 3d-4f transitions is insensitive to optical depth effects
because there are no significant radiative 
decays from the 4f level to the ground term \citep{liu95}; therefore these
transitions are independent of the case assumed. 
In our calculations we have not considered the following lines: lines with
errors higher than 40\%; lines affected
by blends, and the \ion{O}{2} $\lambda$ 4156.54 
line of multiplet 19 because it is presumably blended with an unknown line
\citep{liu00}.
In addition to the 3d-4f transitions, the abundances determined from multiplets
1, 4, 10 and 20 are almost case independent. 
In contrast, multiplets 5, 19 and 25 show strong case sensitivity. This is the reason why
we have adopted as representative of the O$^{++}$/H$^+$ ratio the average 
of values given by multiplets 1, 4, 10, 20 and 3d-4f transitions.

We have detected two 3d-4f transitions belonging to multiplet 55e of \ion{Ne}{2}. For these 
transitions we have used effective recombination coefficients from recent calculations of 
Kissielius \& Storey (unpublished), assuming LS-coupling. 
We have adopted the "sum" value derived from this multiplet:
12+log(Ne$^{++}$/H$^+$)=7.88$^{+0.12}_{-0.16}$  as representative of the Ne$^{++}$
abundance. The values derived from the individual lines as well as the average and the
"sum" value are presented in Table~\ref{neii}.

\section{Total Abundances
\label{abuntot}}

Table ~\ref{totabun} shows the total gaseous abundances of NGC 3576 for
$t^2$=0.00 and the finally adopted ones for $t^2$=0.038. To 
derive the total gaseous abundances, we have to assume a set of ionization 
correction factors, \emph{ICF}'s, to correct for the unseen ionization stages.

The absence of \ion{He}{2} lines
in our spectra indicates that He$^{++}$/H$^+$ is negligible. However, the total
helium abundance has to be corrected for the presence of neutral helium. 
\cite{peim92} determined an \emph{ICF}(He$^0$) = 1+S$^+$/(S-S$^+$), based on the
similarity of the ionization potentials of He$^0$ (24.6 eV) and S$^+$ (23.3 eV). 
With our data, \emph{ICF}(He$^0$) amounts to 1.05 for $t^2$=0.00 and 1.04 for 
$t^2$$>$0.00. He/H is then given by:
\begin{equation}
\frac{N ({\rm He})}{N ({\rm H})} = ICF({\rm He}^0)\times\frac {N({\rm He}^+)}
{N({\rm H}^+)}.                 
\end{equation}

For C we only have direct determinations of C$^{++}$. Therefore, the C abundance is given by:
\begin{equation}
\nonumber
\frac{N(\rm C)}{N(\rm H)} = ICF({\rm C})\times \frac{N({\rm C^{++}})}{N(\rm H^+)}.
\end{equation}
Taking into account the 
similarity between the ionization potentials of C$^{++}$ 
and Ar$^{++}$, and the low Ar$^{3+}$/Ar$^{++}$ ratio obtained, the expected 
C$^{3+}$/C$^{++}$ ratio 
should be negligible. On the other hand, the ionization potential of C$^+$ (24.4 eV) 
is intermediate between 
those of S$^+$ (23.3 eV) and He$^0$ (24.6 eV); therefore we expect 
S$^+$/S$\le$C$^+$/C$\le$He$^0$/He. Moreover, 
to obtain the total He/H ratio we have assumed that S$^+$/S=He$^0$/He.
Therefore, following \citet{peim92} we have 
assumed that S$^+$/S=C$^+$/C, so \emph{ICF}(C)=\emph{ICF}(He$^0$).

To derive the total nitrogen abundance, the usual \emph{ICF}, based on the
similarity 
between the ionization potential of N$^+$ and O$^+$ 
\citep{peim69} is not a good approximation for ionized nebulae with high
degree of ionization. Instead, following \citet{peim92}, we have used the set of
\emph{ICF}s obtained by \citet{mathis91}. We have adopted  the average of the
cool and hot atmosphere results of these authors, which is 0.13 (for $t^2$=0.00)
and 0.15 (for $t^2>$0.00) dex higher than the \emph{ICF} determined using the
standard relation, obtaining a value of 12+log(N/H )=7.87. Alternatively, we
have also derived the total N abundance adding N$^+$/H$^+$
(considering  $t^2$=0.038) and N$^{++}$/H$^+$ determined from permitted lines, obtaining a value 
of 12+log(N/H )=8.07. For comparison, we have used the value of N$^{++}$/H$^+$ calculated by 
\citet{simpson95} from FIR lines, obtaining 12+log(N/H)=7.85 for $t^2$=0.038, in excellent 
agreement with the N abundance derived assuming an \emph{ICF}.
These results indicate, as \citet{tsa03} have pointed out, that \ion{N}{2} lines of
multiplet 3 ---the dominant contributor to our adopted 
N$^{++}$/H$^+$ from RLs--- should be slightly affected by fluorescence effects. 
However, assuming the faint multiplet 48 as representative of the true N$^{++}$/H$^+$, we find 
12+log(N$^{++}$/H$^+$)=7.99$\pm$0.14, which is slightly higher, but consistent within the errors, 
than the adopted N abundance using an \emph{ICF}.

The absence of \ion{He}{2} emission lines in our spectra and the similarity
between the ionization potentials of He$^+$ and O$^{++}$ implies the absence of
O$^{3+}$. We have therefore assumed that:
\begin{equation}
\nonumber
\frac{N(\rm O)}{N(\rm H)}=\frac{N({\rm O^{+}+O^{++}})}{N({\rm H^+})}.
\end{equation}

The only measurable CELs of Ne in the optical region are those of Ne$^{++}$. The
ionization potential of this ion is very high (63.4 eV) and we do not expect a
significant fraction of Ne$^{3+}$. However, Ne$^+$ should be important. Usually,
the \emph{ICF}(Ne$^{++}$) for nebulae with high ionization degree
\citep{peim69} is given by:
\begin{eqnarray}
\nonumber
\frac{N({\rm Ne})}{N({\rm H})} & = &
             ICF({\rm Ne})\times       
             \left[ \frac{N({\rm Ne}^{++})}{N({\rm H}^+)} \right] \\
                               & = & 
             \left[ \frac{N({\rm O})}{N({\rm O}^{++})} \right] \times
             \left[ \frac{N({\rm Ne}^{++})}{N({\rm H}^+)} \right].
\end{eqnarray}

In the case of NGC 3576, this \emph{ICF} gives a correction of 1.50 for $t^2$=0.038, which
implies Ne$^+$/Ne$^{++}$=0.50 and a 12+log(Ne/H)=8.09. On the other hand, \citet{simpson95} have 
observed IR
[\ion{Ne}{2}] lines, obtaining Ne$^+$/Ne$^{++}$=1.0, but in a different zone of NGC
3576 of much lower ionization degree; adopting this value, we obtain a total Ne abundance of
12+log(Ne/H)=8.26, which is 0.17 dex higher than that obtained using the above
equation. On the other hand, \citet{tsa03} obtained Ne$^+$/Ne$^{++}$=0.51 for
the same zone of the nebula, using the same
\emph{ICF}, so we have adopted Ne$^+$/Ne$^{++}$=0.50 to determine the Ne/H ratio.

We have measured CELs of two ionization stages of S, giving S$^+$/S$^{++}$=0.04.
An ionization correction
factor, \emph{ICF}(S), to take into account the presence of S$^{3+}$, has to be
considered. We have adopted the following relation from \citet{sta78}: 
\begin{equation}
\nonumber
ICF(S)=\left[1-\left(1-\frac{N({\rm O^+})}{N(\rm O)}\right)^3\right]^{-1/3},
\end{equation}
and
\begin{equation}
\nonumber
\frac{N(\rm S)}{N(\rm H)} = ICF({\rm S})\times \frac{N({\rm S^{+}+ S^{++}})}{N(\rm H^+)},
\end{equation}
which is based in photoionization models of \ion{H}{2} regions; using this
relation we derived a value of \emph{ICF}(S)=1.1. 
On the other hand, based in the correlation between 
N$^{++}$/N$^+$ $vs.$ S$^{3+}$/S$^{++}$ from ISO observations of compact \ion{H}{2}
regions obtained by \citet{martin02} we estimate an \emph{ICF}(S)=1.2, 
which is in good agreement with that obtained from the above equation. 

We have measured lines of all possible ionization stages of chlorine:
Cl$^+$, Cl$^{++}$ and Cl$^{3+}$. We have derived 
the Cl/H ratio adding the three ionic abundance determinations available for
this element.
However, as we discussed in \S~\ref{cels}, Cl$^+$ atomic data are probably not reliable.
So, alternatively, to take into account the Cl$^+$ fraction, we have adopted the relation by
\citet{peim77}: \emph{ICF}(Cl)=1/(1-S$^+$/S). With our data an \emph{ICF}(Cl)=1.04 is
derived. With this \emph{ICF}, for $t^2$$>$0.00, Cl abundance is 0.02 dex lower than taking 
into account the Cl$^+$/H$^+$ ratio, showing that Cl$^+$ is in fact only a small fraction of the total 
amount of Cl.

For argon we have determinations of Ar$^{++}$ and Ar$^{3+}$. We obtain
Ar$^{3+}$/Ar$^{++}$=0.007, indicating that most Ar is in the form of Ar$^{++}$.
However, some contribution of Ar$^{+}$ is expected. \citet{martin02} have obtained a correlation
between N$^{++}$/N$^+$ $vs.$ Ar$^{++}$/Ar$^{+}$ from ISO observations of compact \ion{H}{2}
regions; using that result we estimate an \emph{ICF}(Ar)=1.1.

Finally, we have measured lines of the three main stages of ionization of iron: Fe$^{+}$,
Fe$^{++}$ and Fe$^{3+}$. \citet{rodri99} has shown evidences for the existence of
fluorescence excitation in the formation process of the observed [\ion{Fe}{2}] lines, so the
determination of the Fe$^+$/H$^+$ ratio is not reliable. On the other hand we have obtained the
Fe$^{3+}$/H$^+$ ratio from the [\ion{Fe}{4}]$\lambda6739.8$ line which has an uncertainty of about 
50\%. In Table~\ref{totabun}, two values are given for the Fe abundance. The first one has
been derived from [\ion{Fe}{3}] and the \emph{ICF} of \citet{rodri04} to take into account 
the fractions of Fe$^+$ and, mainly, Fe$^{3+}$ in the nebula:  
\begin{equation}
\nonumber
\frac{N(\rm Fe)}{N(\rm H)} = \left[\frac{N(\rm O^{+})}{N(\rm O^{++})}\right]^{0.09}\times 
\frac{N(\rm Fe^{++})}{N(\rm O^{+})}\times \frac{N(\rm O)}{N(\rm H)}.
\end{equation}
The second value for the Fe abundance is just the sum of the derived ionic abundances, 
taking into account only Fe$^{++}$ and Fe$^{3+}$ --the contribution of Fe$^+$ should be 
very small (see \S~\ref{cels}). For the handful of objects where [\ion{Fe}{4}] emission has been previously 
measured \citep[see][and references therein]{rodri03} the Fe abundances based
on the sum of the ionic abundances are systematically lower, by factors 2--4, than the
the total abundances implied by Fe$^{++}$ and the above \emph{ICF}. In our case, there are
no differences in the abundances derived from both methods for $t^2$=0.00, and for 
$t^2$$>$0.00 the sum of the
ionic abundances is only a factor of 1.3 lower than the Fe$^{++}$+\emph{ICF} abundance.
This fact could be due either to the lower degree of ionization shown by NGC 3576 respect 
to the other objects \citep[see][]{rodri03, rodri04} or to our possible overestimation of 
the intensity of the extremely weak [\ion{Fe}{4}] $\lambda$6739.8 line.

\section{Discussion
\label{discus}}

In Table~\ref{comparison} we compare the gaseous abundances of NGC 3576 with those
derived by \citet{simpson95} (FIR), \citet{girardi97} (optical CELs) and \citet{tsa03} (optical 
RLs and CELs). 
It can be seen that our
values are in very good agreement with those in common with \citet{tsa03} and rather
similar to the values found by \citet{simpson95}. In contrast, the abundances
differ to the values obtained by \citet{girardi97}, probably because their slit positions are
quite far from ours.
The main differences between our results and those of \citet{tsa03} are in the total
abundances. The different set of \emph{ICF} scheme used, could explain those differences.

To compare the NGC 3576 abundances with those of the Sun, it is necessary
to estimate the fraction of heavy elements embedded in dust grains. We have
assumed that the fraction of heavy elements trapped in dust is the same for NGC 3576 and Orion;
therefore, following EPTE we have added 0.10 dex, 0.08 dex, and 1.37 dex to the gaseous C, O and
Fe abundances, respectively. For N, S, and Cl, no dust correction was applied since they are not
significantly depleted in the neutral ISM \citep{sav96}. For He, Ne, and Ar, no correction was
applied since they are noble gases. 

For the Sun: He comes from \citet{chr98}, C and N 
from \citet{asp03}, O, Ne, and Ar from \citet{asp04}, and S, Cl, and Fe from 
\citet{gre98}.

In Table~\ref{solar} we compare NGC 3576 gas+dust abundances 
with the solar values. We expect a higher O/H value of about 0.15 dex in NGC 3576 than in the
Sun in excellent agreement with the observed value. Our estimate is based on the following considerations:
$i$) from the chemical evolution models for the Galaxy 
\citep{car03,ake04} it is found that the O/H ratio in the interstellar medium
at the solar galactocentric distance has increased by 0.12 dex since the Sun was formed,
$ii$) there is a galactocentric difference of 0.6 kpc in the distance between the Sun and NGC 3576,
and $iii$) the O/H gradient amounts to -0.061 dex kpc$^{-1}$ (see below). Based on the same 
considerations a very good agreement is also found for the excesses obtained 
for Ne and S, the relatively large difference in the Ar/H value is probably due in part
to the uncertain $ICF$ we have used.

\begin{figure}
\begin{center}
\epsscale{1.0}
\plotone{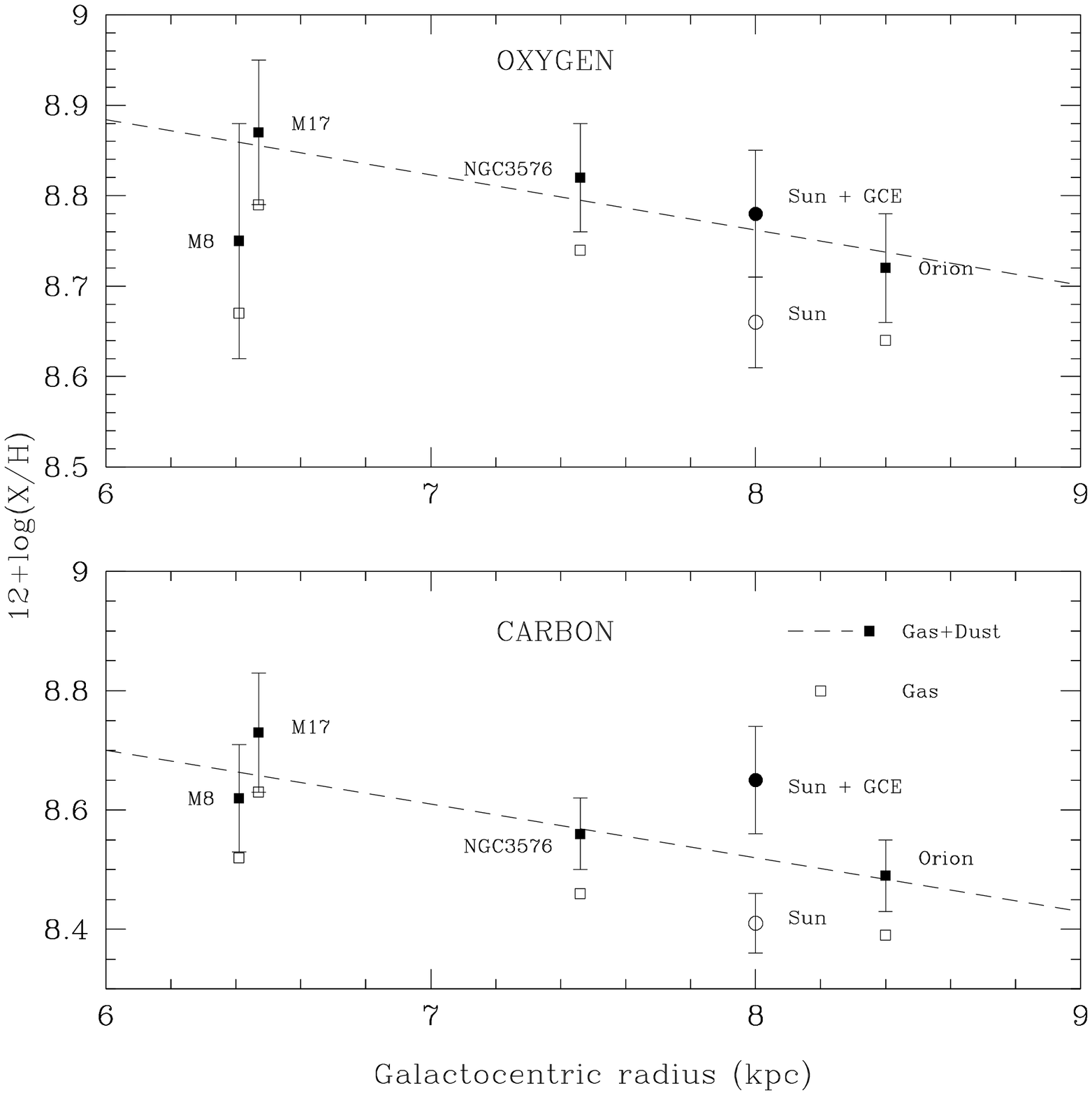}
\figcaption{Galactic O and C radial abundance gradients from \ion{H}{2} region  
abundances determined from recombination lines. Filled squares are dust+gas
abundances, derived applying the corrections proposed by EPTE. Open
squares are the gas-phase abundances. Both sets of data have similar 
error bars, which are only indicated for the filled squares. Abundance data 
for the Orion nebula, M8 and M17 have been taken from EPTE, EPTGR and
\citet{EPTG}, 
respectively. Galactocentric distances have been taken from \citet{russ03}. 
Open circles are the Solar abundances given by \citet{asp03,asp04}. Filled circles are
the values expected for the interstellar medium at the solar Galactocentric distance,
based on the solar values and models for Galactic chemical evolution, GCE \citep{car03,ake04}.
The broken lines represent the correlation found for the \ion{H}{2} regions dust+gas abundances. 
\label{grad}}
\end{center}
\end{figure}

The results of this work, along with those of EPTE, EPTGR and Esteban et al. (1999b) for Orion, M8 
and M17 make possible to present an approach to gas phase abundance
gradients in our Galaxy based exclusively on recombination lines. Figure
~\ref{grad} shows the C/H and O/H abundances derived for these objects. The galactocentric 
distances have been obtained from the complete survey of \citet{russ03} of 
star-forming complexes in our galaxy, using stellar distances to derive their galactocentric 
radius, and adopting a solar galactocentric radius of 8.0 kpc. 
We found a gradient of -0.061 dex kpc$^{-1}$ for O/H, which is somewhat higher 
than the values obtained by \citet{EPTG} and \citet{dehar00}, which are -0.049 and -0.040 dex
kpc$^{-1}$ respectively, and also somewhat higher than the value found for M101 from O RLs by
\citet{este02b}, which is -0.038 dex kpc$^{-1}$. On the other hand, the gradient we derive for 
C/H is -0.090 dex kpc$^{-1}$, which is very similar to the Galactic one derived by 
\citet{EPTG} and revised by \citet{este02b} of -0.086 dex kpc$^{-1}$. Our value of the C gradient 
is consistent with that obtained by \citet{roll00} for B stars: -0.07$\pm$0.02, assuming LTE 
model atmospheres and LTE line formation. However, the absolute C abundances obtained for nearby B 
stars are systematically much lower than the values obtained for the Sun and G-F stars and \ion{H}{2} 
regions of the solar neighborhood. This could be due to NLTE effects or problems with the C atomic 
model used \citep{herr03}. 

Also in Figure~\ref{grad} we show the solar O/H and C/H values and the values
expected for the interstellar medium at the solar galactocentric distance taking into account
the chemical evolution of the Galaxy. From the models by Carigi \citep{car03,ake04} it is found
that the increase in O/H and C/H of the interstellar medium since the Sun was formed 
amounts to 0.12 dex and 0.24 dex respectively.

The C/O gradient is an important constraint for chemical evolution models and the star formation history 
across the Galactic disk. The bulk of these two elements are, in principle, produced by stellar progenitors 
of different initial mass ranges. We derive a C/O gradient of -0.029 dex kpc$^{-1}$, which is similar to that 
given previously by \citet{este02b}: -0.037 dex kpc$^{-1}$; and not too different to that obtained by 
\citet{smar01} for B stars: -0.05 dex kpc$^{-1}$. 
\citet{garn99} have obtained similar C/O gradients in two external spiral galaxies from C abundances derived 
from UV semi-forbidden lines. 

\section{SUMMARY 
\label{conclu}}

We present echelle spectroscopy in the 3100-10400 \AA\ range for the \ion{H}{2} region NGC 3576 (Gum38a).
We have measured the intensities of 461 emission lines; 170 of them are
permitted lines of heavy elements. This is the most complete list of emission lines obtained for this object 
and one of the largest collections ever taken for a Galactic \ion{H}{2} region. 

We have derived physical conditions of the nebula making use of many different line intensities and continuum ratios. 
The chemical abundances have been derived for a large number of ions and different elements. We find excellent agreement 
between the C$^{++}$/H$^+$ ratio obtained from the brightest \ion{C}{2} RL, $\lambda$4267 \AA\ and others 
corresponding to 3d-4f transitions of this ion. All these transitions are ---in principle--- excited by pure recombination 
and give a precise determination of the C$^{++}$ abundance. We find also a good agreement between the O$^{++}$/H$^+$ ratios derived
from RLs of multiplets 1, 4, 10, 20 and 3d-4f, which are case-independent transitions and produced largely by recombination. 
Alternatively, abundances derived for N$^{++}$ for different multiplets show differences as
high as a factor of 3. These differences probably are due to fluorescence
effects. Finally, we have also determined abundances of O$^+$ and Ne$^{++}$ from RLs for the first time in this object.

We have obtained an average $t^2$=0.038$\pm$0.009 both by comparing the
O$^{++}$ and Ne$^{++}$ ionic abundances derived from CELs to those derived from RLs, and by 
comparing the electron temperatures determined from ratios of CELs to those obtained from the Balmer and Paschen 
continua. It is remarkable that the four individual values obtained are almost coincident. The adopted average value of 
$t^2$ has been used to correct the ionic abundances determined from CELs.

We have estimated the C/H, O/H, and C/O Galactic radial abundance gradients making only use of determinations based on RLs 
of \ion{H}{2} regions, obtaining values of -0.090, -0.061, and -0.029, respectively. These estimation is based in four 
objects covering a rather narrow interval of galactocentric distances (from 6 to 9 kpc). 

We would like to thank R. Kisielius and P. J. Storey for providing us with their  latest calculations of effective 
recombination coefficients for Ne, D.P. Smits for providing us unpublished atomic calculations for He, and L. Carigi 
for testing our results with her chemical evolution models. We would also like to thank an
anonymous referee for his/her valuable comments. This work has been partially funded by the Spanish 
Ministerio de Ciencia y Tecnolog\'{\i}a (MCyT) under project AYA2001-0436. MP received partial support from DGAPA UNAM (grant IN
114601). MTR received partial support from FONDAP(15010003), a Guggenheim Fellowship and
Fondecyt(1010404). MR acknowledges support from Mexican CONACYT project J37680-E.

\clearpage

\begin{deluxetable}{ccc} 
\tabletypesize{\scriptsize}
\tablecaption{Log of observations.
\label{tobs}}
\tablewidth{0pt}
\tablehead{
\colhead{Object} & 
\colhead{$\Delta\lambda$ (\AA)} &
\colhead{Exp. Time (s)}} 
\startdata
NGC 3576& B1: 3000--3900& 60, 3 $\times$ 600 \\
{\tt "}& B2: 3800--5000& 120, 3 $\times$ 1800 \\
{\tt "}& R1: 4700--6400& 60, 3 $\times$ 600 \\
{\tt "}& R2: 6300--10400& 120, 3 $\times$ 1800 \\
\enddata
\end{deluxetable}

\begin{deluxetable}{c@{\hspace{5pt}}c@{\hspace{5pt}}cccccc} 
\tabletypesize{\scriptsize}
\tablecaption{Observed and reddening corrected line ratios (F(H$\beta$) = 100) and line 
identifications.
\label{lineid}}
\tablewidth{0pt}
\tablehead{
\colhead{$\lambda_0$ (\AA)} &
\colhead{Ion} &
\colhead{Mult.} &  
\colhead{$\lambda$ (\AA)} &
\colhead{$F(\lambda)$\tablenotemark{a}} &
\colhead{$I(\lambda)$\tablenotemark{b}} &
\colhead{Err(\%)} &
\colhead{notes}} 
\startdata
3187.84 & He I & 3 & 3187.57 & 0.865 & 2.837& 7& \\
3354.55 & He I & 8 & 3354.31 & 0.076 & 0.222& 17& \\
3447.59 & He I & 7 & 3447.40 & 0.131 & 0.311& 12& \\
3478.97 & He I & 48 & 3478.79 & 0.035 & 0.083& 27& \\
 & ? & & 3485.20 & 0.047 & 0.110 & 22& \\
3487.73 & He I & 42 & 3487.54 & 0.047 & 0.110& 22& \\
3498.66 & He I & 40 & 3498.41 & 0.075 & 0.173& 16& \\
3512.52 & He I & 38 & 3512.35 & 0.069 & 0.160& 17& \\
3530.50 & He I & 36 & 3530.28 & 0.099 & 0.228& 14& \\
3554.42 & He I & 34 & 3554.17 & 0.133 & 0.300& 12& \\
3587.28 & He I & 32 & 3587.05 & 0.146 & 0.325& 11& \\
3613.64 & He I & 6 & 3613.42 & 0.212 & 0.467& 9& \\
3634.25 & He I & 28 & 3634.05 & 0.233 & 0.507& 8& \\
3657.27 & H I & H36 & 3657.02 & 0.034 & 0.073& 27& \\
3657.92 & H I & H35 & 3657.67 & 0.030 & 0.065& 30& \\
3658.64 & H I & H34 & 3658.42 & 0.031 & 0.066& 30& \\
3659.42 & H I & H33 & 3659.19 & 0.051 & 0.109& 21& \\
3660.28 & H I & H32 & 3660.05 & 0.077 & 0.166& 16& \\
3661.22 & H I & H31 & 3661.02 & 0.097 & 0.209& 14& \\
3662.26 & H I & H30 & 3661.98 & 0.092 & 0.199& 14& \\
3663.40 & H I & H29 & 3663.18 & 0.123 & 0.263& 12& \\
3664.68 & H I & H28 & 3664.48 & 0.138 & 0.296& 11& \\
3666.10 & H I & H27 & 3665.89 & 0.153 & 0.327& 10& \\
3667.68 & H I & H26 & 3667.45 & 0.177 & 0.379& 10& \\
3669.47 & H I & H25 & 3669.25 & 0.193 & 0.413& 9& \\
3671.48 & H I & H24 & 3671.27 & 0.221 & 0.473& 8& \\
3673.76 & H I & H23 & 3673.56 & 0.259 & 0.552& 8& \\
3676.37 & H I & H22 & 3676.15 & 0.279 & 0.593& 8& \\
3679.36 & H I & H21 & 3679.14 & 0.320 & 0.670& 7& \\
3682.81 & H I & H20 & 3682.58 & 0.352 & 0.747& 7& \\
3686.83 & H I & H19 & 3686.62 & 0.413 & 0.875& 6& \\
3691.56 & H I & H18 & 3691.34 & 0.478 & 1.009& 6& \\
3697.15 & H I & H17 & 3696.93 & 0.577 & 1.215& 5& \\
3703.86 & H I & H16 & 3703.64 & 0.647 & 1.357& 5& \\
3705.04 & He I & 25 & 3704.79 & 0.346 & 0.726& 7& \\
3711.97 & H I & H15 & 3711.76 & 0.763 & 1.593& 5& \\
$\left.\matrix{3721.83\cr 3721.94}\right.$ &
$\left.\matrix{\rm[S\thinspace III]\cr\rm H\thinspace I}\right.$ &
$\left.\matrix{\rm 2F\cr\rm H14}\right\}$ & 
3721.62 & 1.588 & 3.301&  4& \\
3726.03 &[O II] & 1F & 3725.85 & 37.793 & 78.353& 4& \\
3728.82 &[O II] & 1F & 3728.59 & 26.222 & 54.282& 4& \\
3734.37 & H I & H13 & 3734.15 & 1.259 & 2.599& 4& \\
3750.15 & H I & H12 & 3749.93 & 1.550 & 3.171& 4& \\
3770.63 & H I & H11 & 3770.41 & 1.931 & 3.905& 4& \\
3784.89 & He I & 64 & 3784.69 & 0.028 & 0.057& 14& \\
3797.90 & H I & H10 & 3797.68 & 2.763 & 5.528& 4& \\
3805.78 & He I & 63 & 3805.51 & 0.027 & 0.053& 15& \\
3819.61 & He I & 22 & 3819.40 & 0.596 & 1.171& 4&\\
3833.57 & He I & 62 & 3833.28 & 0.039 & 0.076& 11&\\
3835.39 & H I & H9 & 3835.16 & 3.840 & 7.487& 3& \\
3838.37 & N II & 30 & 3837.95 & 0.048 & 0.093& 10& \\
$\left.\matrix{3856.02\cr 3856.13}\right.$ &
$\left.\matrix{\rm Si\thinspace II \cr\rm O\thinspace II}\right.$ &
$\left.\matrix{\rm 1\cr\rm 12}\right\}$ & 
3855.79 & 0.112 & 0.216& 6& \\
3862.59 & Si II & 1 & 3862.37 & 0.099 & 0.191& 7& \\
3867.48 & He I & 20 & 3867.29 & 0.086 & 0.164& 7& \\
3868.75 &[Ne III]& 1F & 3868.51 & 11.373 & 21.748& 3& \\
3871.82 & He I & 60 & 3871.56 & 0.079 & 0.150& 7& \\
$\left.\matrix{3888.65\cr 3889.05}\right.$ &
$\left.\matrix{\rm He\thinspace I \cr\rm H\thinspace I}\right.$ &
$\left.\matrix{\rm 2\cr\rm H8}\right\}$ & 
3888.72 & 8.475 & 16.015& 3& \\
 & ? & & 3914.32 & 0.007 & 0.014& 36& \\
3916.38 & N II & & 3916.17 & 0.007 & 0.014& 35&  \\
3918.98 & C II & 4 & 3918.72 & 0.030 & 0.055& 14& \\
3920.68 & C II & 4 & 3920.43 & 0.054 & 0.100& 9& \\
3926.53 & He I & 58 & 3926.33 & 0.065 & 0.120& 8& \\
3964.73 & He I & 5 & 3964.50 & 0.523 & 0.944& 3& \\
3967.46 &[Ne III]& 1F & 3967.23 & 3.456 & 6.229& 3& \\
3970.07 & H I & H7 & 3969.84 & 8.732 & 15.713& 3& \\
3998.76 & S II & 59 & 3998.53 & 0.009 & 0.016& 31& \\
4008.36 & [Fe III] & 4F & 4008.05 & 0.029 & 0.049& 14& f \\
4009.22 & He I & 55 & 4009.01 & 0.093 & 0.159& 7& \\
4023.98 & He I & 54 & 4023.58 & 0.013 & 0.023 & 23& \\
$\left.\matrix{4026.08\cr 4026.21}\right.$ &
$\left.\matrix{\rm N\thinspace II \cr\rm He\thinspace I}\right.$ &

$\left.\matrix{\rm 40\cr\rm 18}\right\}$ & 
4025.97 & 1.241 & 2.112& 3& \\
4032.81 & S II & 59 & 4032.49 & 0.010 & 0.016& 29& \\
4068.60 & [S II] & 1F & 4068.39 & 0.685 & 1.140& 3& \\
$\left.\matrix{4069.62\cr 4069.89}\right.$ &
$\left.\matrix{\rm O\thinspace II \cr\rm O\thinspace II}\right.$ &
$\left.\matrix{\rm 10\cr\rm 10}\right\}$ & 
4069.53 & 0.084 & 0.139& 7& \\
4072.15 & O II & 10 & 4071.93 & 0.051 & 0.085& 9& \\
4075.86 & O II & 10 & 4075.62 & 0.065 & 0.108& 8& \\
4076.35 & [S II] & 1F & 4076.14 & 0.232 & 0.385& 4& \\
4083.90 & O II & 48b & 4083.52 & 0.011 & 0.017& 28& \\
4085.11 & O II & 10 & 4084.90 & 0.015 & 0.024& 22& \\
4087.15 & O II & 48c & 4086.88 & 0.016 & 0.026& 21& \\
4089.29 & O II & 48a & 4089.02 & 0.027 & 0.044& 14& e\\
$\left.\matrix{4097.25\cr 4097.26}\right.$ &
$\left.\matrix{\rm O\thinspace II \cr\rm O\thinspace II}\right.$ &
$\left.\matrix{\rm 20\cr\rm 48b}\right\}$ & 
4097.04 & 0.032 & 0.052& 13& \\
4101.74 & H I & H6 & 4101.50 & 15.369 & 24.681& 3& \\
4110.78 & O II & 20 & 4110.51 & 0.009 & 0.015& 30& \\
4119.22 & O II & 20 & 4119.05 & 0.014 & 0.023& 22& \\
4120.84 & He I & 16 & 4120.58 & 0.143 & 0.232& 5& e\\
4129.32 & O II & 19 & 4129.00 & 0.006 & 0.009& :& \\
4131.72 & Ar II & & 4131.48 & 0.008 & 0.013& 33& g \\
4132.80 & O II & 19 & 4132.57 & 0.022 & 0.035& 17& \\
4143.76 & He I & 53 & 4143.52 & 0.178 & 0.284& 4& \\
$\left.\matrix{4145.91\cr 4146.09}\right.$ &
$\left.\matrix{\rm O\thinspace II \cr\rm O\thinspace II}\right.$ &
$\left.\matrix{\rm 106\cr\rm 106}\right\}$ & 
4145.64 & 0.012 & 0.019& 26& \\
4153.30 & O II & 19 & 4153.05 & 0.027 & 0.043& 14& \\
4156.54 & O II & 19 & 4156.03 & 0.018 & 0.028& 19& f\\
$\left.\matrix{4168.97\cr 4169.22}\right.$ &
$\left.\matrix{\rm He\thinspace I \cr\rm O\thinspace II}\right.$ &
$\left.\matrix{\rm 52\cr\rm 19}\right\}$ & 
4168.76 & 0.041 & 0.064& 11& \\
4185.45 & O II & 36 & 4185.22 & 0.022 & 0.034& 17& \\
4189.79 & O II & 36 & 4189.49 & 0.017 & 0.026& 20&  \\
4201.35 & N II & 49 & 4201.01 & 0.009 & 0.014& 30& g \\
4236.91 & N II & 48 & 4236.64 & 0.007 & 0.011& 36& \\
4241.78 & N II & 48 & 4241.53 & 0.009 & 0.014& 31&  \\
4242.50 & N II & 48 & 4242.29 & 0.006 & 0.010& 39& \\
4243.97 & [Fe II]& 21F & 4243.76 & 0.019 & 0.028& 18& \\
4267.15 & C II & 6 & 4266.91 & 0.199 & 0.295& 4& \\
4275.55 & O II & 67a & 4275.33 & 0.013 & 0.019& 24& e \\
$\left.\matrix{4276.75\cr 4276.83}\right.$ &
$\left.\matrix{\rm O\thinspace II \cr\rm [Fe\thinspace II]}\right.$ &
$\left.\matrix{\rm 67b\cr\rm 21F}\right\}$ & 
4276.51 & 0.038 & 0.055& 11& e \\
4285.69 & O II & 78b & 4285.28 & 0.010 & 0.015& 29& \\
4287.39 & [Fe II]& 7F & 4287.19 & 0.055 & 0.081& 9& \\
$\left.\matrix{4303.61\cr 4303.82}\right.$ &
$\left.\matrix{\rm O\thinspace II \cr\rm O\thinspace II}\right.$ &
$\left.\matrix{\rm 65a\cr\rm 53a}\right\}$ & 
4303.56 & 0.021 & 0.031& 17& \\
4317.14 & O II & 2 & 4316.86 & 0.018 & 0.026& 19& \\
4319.63 & O II & 2 & 4319.35 & 0.014 & 0.019& 23& \\
4326.40 & O I & & 4326.14 & 0.019 & 0.027& 18& \\
4332.71 & O II & 65b & 4332.45 & 0.009 & 0.013& :&  \\
4336.79 & [Cr II]& & 4336.53 & 0.040 & 0.056& 11& \\
4340.47 & H I & H5 & 4340.20 & 32.386 & 45.921& 2\\
$\left.\matrix{4345.55\cr 4345.56}\right.$ &
$\left.\matrix{\rm O\thinspace II \cr\rm O\thinspace II}\right.$ &
$\left.\matrix{\rm 65c\cr\rm 2}\right\}$ & 
4345.22 & 0.040 & 0.057& 11& \\
4349.43 & O II & 2 & 4349.18 & 0.047 & 0.067& 10& \\
4359.33 & [Fe II]& 7F & 4359.13 & 0.037 & 0.051& 12&  \\
4363.21 & [O III]& 2F & 4362.95 & 0.915 & 1.279& 2& \\
4366.89 & O II & 2 & 4366.62 & 0.029 & 0.040& 14& \\
4368.25 & O I & 5 & 4368.10 & 0.050 & 0.069& 9& \\
4372.43 & [Fe II]& 21F & 4372.24 & 0.006 & 0.009& :& \\
4387.93 & He I & 51 & 4387.67 & 0.404 & 0.555& 3& \\
4391.99 & Ne II & 55e & 4391.66 & 0.013 & 0.018& :& \\
4409.30 & Ne II & 55e & 4408.89 & 0.014 & 0.024& :& \\
4413.78 & [Fe II]& 7F & 4413.57 & 0.029 & 0.050& 14& \\
4414.90 & O II & 5 & 4414.64 & 0.017 & 0.029& 20& \\
4416.27 & [Fe II]& 6F & 4416.07 & 0.039 & 0.067& 11& \\
4416.97 & O II & 5 & 4416.71 & 0.018 & 0.031& 19& \\
4437.55 & He I & 50 & 4437.29 & 0.048 & 0.080& 10& \\
$\left.\matrix{4452.10\cr 4452.37}\right.$ &
$\left.\matrix{\rm [Fe\thinspace II] \cr\rm O\thinspace II}\right.$ &
$\left.\matrix{\rm 7F\cr\rm 5}\right\}$ & 
4451.91 & 0.021 & 0.035& 17& \\
4471.09 & He I & 14 & 4471.24 & 3.435 & 5.503& 2& \\
4474.90 & [Fe II]& 7F & 4474.72 & 0.015 & 0.024& 22& \\
$\left.\matrix{4491.07 \cr 4491.23 }\right.$ &
$\left.\matrix{\rm C\thinspace II \cr\rm O\thinspace II}\right.$ &
$\left.\matrix{\rm \cr\rm 86a}\right\}$ & 
4491.03 & 0.014 & 0.022& 22& \\
4562.60 & Mg I] & 1 & 4562.17 & 0.0095 & 0.0135& 30& \\
4571.10 & Mg I]& 1 & 4570.87 & 0.014 & 0.019& 23& \\
4590.97 & O II & 15 & 4590.67 & 0.021 & 0.029& 17& \\
$\left.\matrix{4595.95\cr 4596.18}\right.$ &
$\left.\matrix{\rm O\thinspace II \cr\rm O\thinspace II}\right.$ &
$\left.\matrix{\rm 15\cr\rm 15}\right\}$ & 
4595.92 & 0.018 & 0.025& 19&\\
4601.48 & N II & 5 & 4601.25 & 0.010 & 0.013& 29& \\
4602.13 & O II & 92b & 4601.72 & 0.004 & 0.006& :& \\
$\left.\matrix{4607.06\cr 4607.16}\right.$ &
$\left.\matrix{\rm [Fe\thinspace III] \cr\rm N\thinspace II}\right.$ &
$\left.\matrix{\rm 3F\cr\rm 5}\right\}$ & 
4606.85 & 0.035 & 0.047& 12& \\
4609.44 & O II & 92a & 4609.16 & 0.018 & 0.024& 19& \\
$\left.\matrix{4613.68\cr 4613.87}\right.$ &
$\left.\matrix{\rm O\thinspace II \cr\rm N\thinspace II}\right.$ &
$\left.\matrix{\rm 92b\cr\rm 5}\right\}$ & 
4613.58 & 0.008 & 0.010& 35& \\
4621.39 & N II & 5 & 4621.12 & 0.017 & 0.023& 20& \\
4624.11 & S II & & 4623.85 & 0.005 & 0.006& :& g \\
4630.54 & N II & 5 & 4630.26 & 0.043 & 0.055& 10& \\
4634.14 & N III & 2 & 4633.82 & 0.012 & 0.015& 26& \\
4638.86 & O II & 1 & 4638.57 & 0.057 & 0.074& 8& \\
4640.64 & N III & 2 & 4640.38 & 0.027 & 0.034& 14& \\
4641.81 & O II & 1 & 4641.53 & 0.103 & 0.132& 6& \\
4643.06 & N II & 5 & 4642.78 & 0.021 & 0.027& 17& \\
4649.13 & O II & 1 & 4648.86 & 0.114 & 0.145& 5& \\
4650.84 & O II & 1 & 4650.54 & 0.055 & 0.069& 9& \\
4658.10 &[Fe III]& 3F & 4657.87 & 0.438 & 0.552& 2& \\
4661.63 & O II & 1 & 4661.30 & 0.072 & 0.090& 7& e\\
4667.01 &[Fe III]& 3F & 4666.70 & 0.024 & 0.030& 16& f \\
4673.73 & O II & 1 & 4673.39 & 0.011 & 0.013& 27& \\
4676.24 & O II & 1 & 4675.92 & 0.032 & 0.040& 13& \\
4696.36 & O II & 1 & 4696.04 & 0.006 & 0.007& :& \\
4699.21 & O II & 25 & 4698.80 & 0.006 & 0.007& :& \\
4705.35 & O II & 25 & 4705.09 & 0.007 & 0.008& 38& \\
4701.53 &[Fe III]& 3F & 4701.33 & 0.121 & 0.144& 5& \\
4711.37 &[Ar IV] & 1F & 4711.12 & 0.042 & 0.050& 10& \\
4713.14 & He I & 12 & 4712.90 & 0.525 & 0.620& 2& \\
4733.91 &[Fe III]& 3F & 4733.65 & 0.048 & 0.055& 10& \\
4740.16 &[Ar IV] & 1F & 4739.95 & 0.045 & 0.051& 10& \\
4752.96 & O II & 94 & 4752.60 & 0.007 & 0.008& 35& \\
4754.69 &[Fe III]& 3F & 4754.50 & 0.082 & 0.092& 7& \\
4769.43 &[Fe III]& 3F & 4769.23 & 0.048 & 0.053& 10& \\
4777.68 &[Fe III]& 3F & 4777.47 & 0.024 & 0.027& 15& \\
4779.71 & N II & 20 & 4779.34 & 0.008 & 0.009& 32& \\
4788.13 & N II & 20 & 4787.73 & 0.016 & 0.017& 20& \\
4792.01 & S II & 46 & 4791.76 & 0.008 & 0.009& 33& \\
4802.23 & C II &  & 4802.16 & 0.009 & 0.010& 30& \\
4803.29 & N II & 20 & 4802.99 & 0.018 & 0.019& 19& \\
4814.55 &[Fe II] & 20F & 4814.31 & 0.031 & 0.033& 21& \\
4815.51 & S II & 9 & 4815.26 & 0.016 & 0.017& 21& \\
4861.33 & H I & H4 & 4861.09 &100.000 & 100.000 & 0.7& \\
4881.00 &[Fe III]& 2F & 4880.83 & 0.214 & 0.209& 5\\
$\left.\matrix{4889.63\cr 4889.70}\right.$ &
$\left.\matrix{\rm [Fe\thinspace II] \cr\rm [Fe\thinspace II]}\right.$ &
$\left.\matrix{\rm 4F\cr\rm 3F}\right\}$ & 
4889.44 & 0.014 & 0.014& 23 & \\
4902.65 & Si II &7.23 & 4902.41 & 0.014 & 0.014& 22& \\
4905.34 &[Fe II] & 20F & 4905.17 & 0.015 & 0.015& 21&  \\
4921.93 & He I & 48 & 4921.69 & 1.258 & 1.183& 2& \\
$\left.\matrix{4924.50\cr 4924.50}\right.$ &
$\left.\matrix{\rm [Fe\thinspace III] \cr\rm O\thinspace II}\right.$ &
$\left.\matrix{\rm 2F\cr\rm 28}\right\}$ & 
4924.32 & 0.045 & 0.042& 10& \\
4930.50 &[Fe III]& 1F & 4930.32 & 0.011 & 0.010& 27& \\
4931.32 &[O III] & 1F & 4930.97 & 0.054 & 0.051& 19& \\
4958.91 &[O III] & 1F & 4958.69 &134.179 &121.335& 0.7& \\
4985.90 &[Fe III]& 2F & 4985.58 & 0.041 & 0.036& 11&  \\
4987.20 &[Fe III]& 2F & 4987.00 & 0.033 & 0.029& 12&  \\
4994.37 & N II & 94 & 4994.16 & 0.033 & 0.029& 28& \\
4996.98 & O II & & 4996.76 & 0.043 & 0.037& 23& g \\
5001.47 & N II & 19 & 5001.15 & 0.042 & 0.037& 24& \\
5006.84 &[O III] & 1F & 5006.66 &408.677 &353.023& 0.7& \\
5011.30 &[Fe III]& 1F & 5011.16 & 0.060 & 0.051& 19& \\
5015.68 & He I & 4 & 5015.47 & 2.645 & 2.266& 2& \\
5035.79 &[Fe II] & 4F & 5035.59 & 0.025 & 0.021& 34& \\
5041.03 & Si II & 5 & 5040.82 & 0.255 & 0.213& 7& \\
5041.98 & O II &23.01 & 5041.78 & 0.012 & 0.010& :& \\
5045.10 & N II & 4 & 5044.79 & 0.029 & 0.024& 30& \\
5047.74 & He I & 47 & 5047.61 & 0.389 & 0.323& 5& e\\
$\left.\matrix{5055.98\cr 5056.31}\right.$ &
$\left.\matrix{\rm Si\thinspace II \cr\rm Si\thinspace II}\right.$ &
$\left.\matrix{\rm 5\cr\rm 5}\right\}$ & 
5055.84 & 0.262 & 0.216& 7& \\
5084.77 &[Fe III]& 1F & 5084.58 & 0.012 & 0.010& :& \\
5111.63 &[Fe II] & 19F & 5111.51 & 0.016 & 0.012& :& \\
5121.83 & C II & & 5121.61 & 0.011 & 0.009& :& \\
5146.70 &[Fe III]& & 5146.49 & 0.014 & 0.011& :& g \\
5158.78 &[Fe II] & 19F & 5158.62 & 0.063 & 0.047& 18& \\
5191.82 &[Ar III]& 3F & 5191.50 & 0.099 & 0.072& 13& \\
5197.90 & [N I] & 1F & --- & --- & --- & ---& c \\
5200.26 & [N I] & 1F & --- & --- & --- & ---& c \\
5261.61 &[Fe II] & 19F & 5261.51 & 0.053 & 0.037& 20&  \\
5270.40 &[Fe III]& 1F & 5270.33 & 0.332 & 0.227& 6& \\
5273.35 &[Fe II] & 18F & 5273.19 & 0.025 & 0.017& 34& \\
5275.12 & O I & 27 & 5275.17 & 0.022 & 0.015& 37& \\
5276.85 & C II & 56 & 5276.55 & 0.015 & 0.010& :& g \\
5299.00 & O I & 26 & 5298.99 & 0.040 & 0.027& 24& \\
5333.65 &[Fe II] & 19F & 5333.47 & 0.015 & 0.010& :& \\
5342.38 & C II & 17.06 & 5342.05 & 0.018 & 0.012& :& \\
5412.00 &[Fe III]& 1F & 5411.95 & 0.037 & 0.023& 26& \\
5423.20 & N I & & 5422.86 & 0.009 & 0.006&  :& g \\
5432.77 & S II & 6 & 5432.54 & 0.022 & 0.013& 37& \\
5453.81 & S II & 6 & 5453.69 & 0.027 & 0.016& 32& \\
$\left.\matrix{5495.70\cr 5495.82}\right.$ &
$\left.\matrix{\rm N\thinspace II \cr\rm [Fe\thinspace II]}\right.$ &
$\left.\matrix{\rm 29\cr\rm 17F}\right\}$ & 
5495.43 & 0.013 & 0.008& 53& \\
5506.87 &[Cr III]& & 5506.52 & 0.008 & 0.004& :&\\
5512.77 & O I & 25 & 5512.60 & 0.029 & 0.014& 31&\\
5517.71 &[Cl III]& 1F & 5517.42 & 0.727 & 0.359& 5&\\
5537.88 &[Cl III]& 1F & 5537.60 & 0.807 & 0.396& 5& \\
5542.58 & S I & & 5542.41 & 0.015 & 0.007& :& g \\
$\left.\matrix{5545.00\cr 5545.15}\right.$ &
$\left.\matrix{\rm N\thinspace I \cr\rm N\thinspace I}\right.$ &
$\left.\matrix{\rm 29\cr\rm 29}\right\}$ & 
5544.89 & 0.020 & 0.010& 39& \\
5551.95 & N II & 63 & 5551.63 & 0.012 & 0.006& :& g \\
5555.03 & O I & 24 & 5554.79 & 0.022 & 0.011& 37& \\
5577.34 & [O I] & 3F & --- & --- & --- & ---& c \\
5581.86 &[Fe II] & 15F & 5581.50 & 0.015 & 0.007& : & g \\
5666.64 & N II & 3 & 5666.32 & 0.045 & 0.021& 23& \\
5676.02 & N II & 3 & 5675.68 & 0.030 & 0.014& 30& \\
5679.56 & N II & 3 & 5679.29 & 0.084 & 0.039& 15& \\
5686.21 & N II & 3 & 5685.90 & 0.011 & 0.005& :& \\
5710.76 & N II & 3 & 5710.52 & 0.011 & 0.005& :& \\
5754.64 & [N II] & 3F & 5754.36 & 0.887 & 0.404& 5& \\
5875.64 & He I & 11 & 5875.34 & 26.060 & 11.373& 4& \\
5907.21 & C II & 44 & 5906.96 & 0.025 & 0.011& 34& \\
5927.82 & N II & 28 & 5927.48 & 0.021 & 0.009& 38& \\
5931.79 & N II & 28 & 5931.47 & 0.047 & 0.020& 22&\\
5940.24 & N II & 28 & 5939.89 & 0.014 & 0.006& :&\\
5941.68 & N II & 28 & 5941.29 & 0.034 & 0.015& 27&\\
5957.56 & Si II & 4 & 5957.29 & 0.053 & 0.023& 21&\\
5958.58 & O I & 23 & 5958.40 & 0.095 & 0.040& 14&  \\
5978.93 & Si II & 4 & 5978.67 & 0.105 & 0.044& 13& \\
6000.20 &[Ni III]& 2F & 5999.94 & 0.016 & 0.007& :& \\
6046.44 & O I & 22 & 6046.19 & 0.095 & 0.039& 14& \\
6151.43 & C II &16.04 & 6151.17 & 0.030 & 0.012& 30& \\
6300.30 & [O I] & 1F & 6300.05 & 1.515 & 0.570 & 5& c \\
6312.10 &[S III] & 3F & 6311.73 & 4.122 & 1.544& 5& \\
6347.11 & Si II & 2 & 6346.77 & 0.387 & 0.143& 7& \\
6363.78 & [O I] & 1F & 6363.53 & 0.556 & 0.205 & 6& c \\
6371.36 & Si II & 2 & 6371.00 & 0.353 & 0.129& 8& \\
6454.80 & C II &17.05F & 6454.54 & 0.011 & 0.004& :&  \\
6462.00 & C II &17.04 & 6461.53 & 0.090 & 0.032& 15&  \\
6527.10 & [N II] & 1F & 6526.87 & 0.021 & 0.007& 38& \\
6548.03 &[N II] & 1F & 6547.76 & 35.114 & 12.107& 5& \\
6562.21 & H I & H3 & 6562.43 &768.576 &263.627& 5& \\
6578.05 & C II & 2 & 6577.64 & 0.731 & 0.249& 6& \\
6583.41 &[N II] & 1F & 6583.12 &112.131 & 38.183& 5& \\
6666.80 &[Ni II] & 2F & 6666.53 & 0.019 & 0.006& :& \\
6678.15 & He I & 46 & 6677.76 & 10.897 & 3.589& 6& \\
6716.47 & [S II] & 2F & 6716.12 & 17.940 & 5.830& 6& \\
6721.39 & O II & 4 & 6720.92 & 0.011 & 0.004& 27& \\
6730.85 &[S II] & 2F & 6730.50 & 22.585 & 7.302& 6& \\
6733.90 & [Cr IV]& $^{4}$F-$^{2}$G & 6733.62 & 0.012 & 0.004& 26& g\\
6739.80 &[Fe IV] &$^2$G--$^2$I & 6739.75 & 0.020 & 0.006& 18&  \\
6744.39 & C II & & 6744.08 & 0.015 & 0.005& 22& \\
6747.50 &[Cr IV] &$^4$F--$^2$G & 6747.21 & 0.012 & 0.004& 26& \\
6755.90 & He I & 1/20 & 6755.49 & 0.013 & 0.004& 24& g \\
6769.61 & N I & & 6769.22 & 0.014 & 0.004& 23& g \\
6785.81 & O II &  & 6785.37 & 0.012 & 0.004& 26& e, g\\
6791.25 & Ne II & & 6790.94 & 0.015 & 0.005& 22& g \\
6813.57 &[Ni II] & 8F & 6813.27 & 0.008 & 0.003& 33& \\
6818.22 & N II & & 6817.97 & 0.008 & 0.003& 33& g \\
6855.88 & He I & 1/12 & & & & & e\\
7002.23 & O I & 21 & 7001.82 & 0.213 & 0.067& 7&  c \\
7062.26 & He I &1/11 & 7062.04 & 0.074 & 0.023& 9&  \\
7065.28 & He I & 10 & 7064.82 & 19.453 & 5.919& 6& \\
7083.00 & Ar I & & 7082.70 & 0.035 & 0.010& 13& g \\
7110.90 &[Cr IV] & & 7110.59 & 0.017 & 0.005& 20& \\
7113.42 & Si II & & 7112.87 & 0.013 & 0.004& 24& g \\
7115.40 & Si I & & 7115.13 & 0.011 & 0.003& 28& g \\
7135.78 &[Ar III]& 1F & 7135.36 & 53.072 & 15.581& 6& \\
7155.14 &[Fe II] & 14F & 7154.83 & 0.126 & 0.036& 8& \\
7160.58 & He I &1/10 & 7160.13 & 0.069 & 0.020& 10&  \\
7231.12 & C II & 3 & 7230.84 & 0.332 & 0.093& 7& \\
7236.19 & C II & 3 & 7235.94 & 0.531 & 0.148& 7& \\
7237.17 & C II & 3 & 7236.80 & 0.067 & 0.019& 10& \\
7254.38 & O I & 20 & 7254.21 & 0.134 & 0.037& 8& \\
7281.35 & He I & 45 & 7280.92 & 2.447 & 0.669& 7& \\
7298.05 & He I & 1/9 & 7297.61 & 0.104 & 0.028& 9&  \\
7318.39 & [O II] & 2F & 7318.63 & 2.358 & 0.633& 7& \\
7319.99 & [O II] & 2F & 7319.74 & 9.400 & 2.522& 7& \\
7329.66 & [O II] & 2F & 7329.31 & 5.540 & 1.479& 7& c \\
7330.73 & [O II] & 2F & 7330.38 & 5.078 & 1.355& 7& \\
7377.83 & [Ni II]& 2F & 7377.53 & 0.202 & 0.053& 8& \\
7388.17 & [Fe II]& 14F & 7387.75 & 0.021 & 0.006& 18& \\
7390.60 & [Cr IV]& 1F & 7390.38 & 0.029 & 0.007& 15& \\
7411.61 & [Ni II]& 2F & 7411.34 & 0.051 & 0.013& 11& \\
7423.64 & N I & 3 & 7423.29 & 0.054 & 0.014& 11& \\
7442.30 & N I & 3 & 7442.01 & 0.109 & 0.028& 9& \\
7452.54 & [Fe II]& 14F & 7452.20 & 0.049 & 0.012& 12& \\
7468.31 & N I & 3 & 7467.99 & 0.168 & 0.042& 8& \\
7477.10 & Si I & & 7476.56  & 0.012 & 0.003& 26 & g \\
7499.18 & He I & 1/8 & 7499.43 & 0.177 & 0.044& 8&  \\
7504.94 & O II & & 7504.51 & 0.015 & 0.004& 22& \\
 & ? & & 7512.83 & 0.034 & 0.008& 14& \\
7519.86 & Si I & & 7519.42 & 0.019 & 0.005& 20& \\
7530.54 & [Cl IV]& 1F& 7529.96  & 0.032 & 0.008& 14 \\
7538.06 & Si I & & 7537.56 & 0.008 & 0.002& 35& g \\
7714.54 & He I &2/15 & 7714.04 & 0.020 & 0.005& 19&  \\
7751.10 &[Ar III]& 2F & 7750.68 & 17.582 & 3.894& 8& c \\
7771.94 & O I & 1 & 7771.38 & 0.051 & 0.011& 12& c \\
7774.17 & O I & 1 & 7773.66 & 0.254 & 0.056& 8& c \\
7816.13 & He I & 1/7 & 7815.68 & 0.321 & 0.069& 8&  \\
7837.76 &Ar II & & 7837.42  & 0.012 & 0.003& 26 & g \\
7875.99 & [P II] & & 7875.46 & 0.038 & 0.008& 14& g  \\
7971.62 & He I &2/11 & 7971.08 & 0.041 & 0.008& 13&  \\
8000.08 & [Cr II]& 1F & 7999.56 & 0.074 & 0.015& 11& \\
8030.69 & Ar II & & 8030.31 & 0.022 & 0.004& 18& g \\
8045.63 & [Cl IV]& 1F & 8045.24 & 0.075 & 0.015& 11& \\
8057. & He I &4/18 & 8057.07 & 0.030 & 0.006& 16& \\
8084. & He I &4/17 & 8083.78 & 0.034 & 0.007&  15&  \\
8094.08 & He I &2/10 & 8093.85 & 0.193 & 0.037& 9&  \\
8116. & He I &4/16 & 8115.91 & 0.041 & 0.008& 13&  \\
8125.30 & [Cr II] & 1F& 8124.91 & 0.033 & 0.006& 15& d \\
8184.85 & N I & 2 & 8184.53 & 0.098 & 0.018& 10& d \\
8188.01 & N I & 2 & 8187.65 & 0.191 & 0.036& 14& d \\
8200.91 & C II & & 8200. & --- & --- & & c \\
8203.85 & He I &4/14 & 8203.35 & 0.071 & 0.013& 11&  \\
8210.72 & N I & 2 & 8210.34 & 0.052 & 0.010& 12& \\
8216.28 & N I & 2 & 8215.96 & 0.236 & 0.044& 9& \\
8245.64 & H I & P42 & 8245.13 & 0.212 & 0.039& 9& \\
8247.73 & H I & P41 & 8247.27 & 0.215 & 0.039& 9& \\
8249.20 & H I & P40 & 8249.48 & 0.222 & 0.041& 9& \\
8252.40 & H I & P39 & 8251.90 & 0.263 & 0.048& 9& \\
8255.02 & H I & P38 & 8254.55 & 0.306 & 0.056& 9& \\
8257.85 & H I & P37 & 8257.45 & 0.228 & 0.042& 9& \\
8260.93 & H I & P36 & 8260.52 & 0.279 & 0.051& 9& \\
8264.28 & H I & P35 & 8263.92 & 0.351 & 0.064& 9& \\
$\left.\matrix{8265.71\cr 8265.71}\right.$ &
$\left.\matrix{\rm He\thinspace I \cr\rm He\thinspace I}\right.$ &
$\left.\matrix{\rm 4/13\cr\rm 2/9}\right\}$ & 
8265.29 & 0.088 & 0.016&  11&  \\
8267.94 & H I & P34 & 8267.46 & 0.395 & 0.072& 9& \\
8271.93 & H I & P33 & 8271.40 & 0.423 & 0.077& 9& \\
8276.31 & H I & P32 & 8275.83 & 0.450 & 0.082& 9& \\
8281.12 & H I & P31 & 8280.50 & 0.376 & 0.068& 9& c, d\\
8286.43 & H I & P30 & 8285.86 & 0.441 & 0.080& 9& \\
8292.31 & H I & P29 & 8291.78 & 0.652 & 0.118& 9& \\
8298.83 & H I & P28 & 8298.22 & 0.492 & 0.089& 9& \\
8306.11 & H I & P27 & 8305.77 & 0.432 & 0.078& 9& \\
8314.26 & H I & P26 & 8313.75 & 0.683 & 0.122& 9& \\
8323.42 & H I & P25 & 8322.94 & 0.792 & 0.141& 9& \\
8333.78 & H I & P24 & 8333.29 & 0.856 & 0.152& 9& \\
8342.33 & He I &4/12 & 8341.85 & 0.111 & 0.020&  10&  \\
8359.00 & H I & P22 & 8358.51 & 1.125 & 0.198& 9& \\
8361.67 & He I & 1/6 & 8361.22 & 0.624 & 0.110& 9&  \\
8374.48 & H I & P21 & 8373.98 & 1.134 & 0.199& 9& \\
8376. & He I &6/20 & 8375.95 & 0.068 & 0.012& 11&  \\
8388.00 & Ar I & & 8387.35  & 0.024 & 0.004& 18 & g \\
8392.40 & H I & P20 & 8391.89 & 1.459 & 0.254& 9& \\
8397. & He I &6/19 & 8396.86 & 0.057 & 0.010& 12&  \\
8413.32 & H I & P19 & 8412.82 & 1.644 & 0.284& 9& \\
8422. & He I &6/18 & 8421.45 & 0.062 & 0.011& 12&  \\
8424. & He I &7/18 & 8423.91 & 0.039 & 0.007& 14&  \\
8433.85 & [Cl III] & 3F & 8432.99 & 0.048 & 0.008 & 13& g \\
8437.96 & H I & P18 & 8437.46 & 1.892 & 0.325& 9& \\
$\left.\matrix{8444.34\cr 8444.34}\right.$ &
$\left.\matrix{\rm He\thinspace I \cr\rm N\thinspace III}\right.$ &
$\left.\matrix{\rm 4/11\cr\rm 267}\right\}$ & 
8444.00 & 0.161 & 0.028& 10& c \\
8446.48 & O I & 4 & 8446.12 & 3.725 & 0.638& 9& c\\
8451.00 & He I &6/17 & 8450.70 & 0.080 & 0.014& 11&  \\
8467.25 & H I & P17 & 8466.76 & 2.194 & 0.373& 9& \\
8480.90 &[Cl III]& 3F & 8480.36 & 0.049 & 0.008& 13& \\
8486. & He I &6/16 & 8485.80 & 0.102 & 0.017& 10&  \\
8488. & He I &7/16 & 8488.26 & 0.040 & 0.007& 14&  \\
8500.00 &[Cl III]& 3F & 8499.35 & 0.104 & 0.018& 10& \\
8502.48 & H I & P16 & 8501.98 & 2.690 & 0.452& 9& \\
8518.04 & He I & 2/8 & 8517.52 & 0.070 & 0.012& 11&  \\
8528.99 & He I & 6/15 & 8528.54 & 0.136 & 0.023& 10&  \\
8531.48 & He I & 7/15 & 8531.08 & 0.051 & 0.009& 13&  \\
8665.02 & H I & P13 & 8664.48 & 5.302 & 0.847& 9&\\
8680.28 & N I & 1 & 8680.07 & 0.265 & 0.042& 10& \\
8683.40 & N I & 1 & 8683.02 & 0.179 & 0.028& 10& \\
8686.15 & N I & 1 & 8685.87 & 0.183 & 0.029& 10& \\
8703.25 & N I & 1 & 8702.87 & 0.123 & 0.019& 10& \\
8711.70 & N I & 1 & 8711.32 & 0.144 & 0.023& 10& \\
8718.84 & N I & 1 & 8718.46 & 0.075 & 0.012& 12& \\
8727.13 & [C I] & 3F & 8726.85 & 0.162 & 0.025& 10& c \\
$\left.\matrix{8728.90\cr 8728.90}\right.$ &
$\left.\matrix{\rm [Fe\thinspace III] \cr\rm N\thinspace I}\right.$ &
$\left.\matrix{\rm 8F\cr\rm 28}\right\}$ & 
8728.76 & 0.063 & 0.010& 12& \\
8733.43 & He I & 6/12 & 8732.92 & 0.239 & 0.037& 10&  \\
8736.04 & He I & 7/12 & 8735.51 & 0.078 & 0.012& 11&  \\
8737.83 & Ar II & & 8737.80 & 0.016 & 0.003& 23& g\\
8739.97 & He I & 5/12 & 8739.54 & 0.023 & 0.004& 18&  \\
8750.47 & H I & P12 & 8749.96 & 7.051 & 1.098& 9& \\
8776.77 & He I & 4/9 & 8776.44 & 1.185 & 0.183& 9&  \\
8816.82 & He I &10/12 & 8816.13 & 0.044 & 0.007& 14&  \\
8829.40 & [S III]& 3F & 8829.14 & 0.088 & 0.013& 11& \\
8845.38 & He I & 6/11 & 8844.94 & 0.401 & 0.061& 10&  \\
8848.05 & He I & 7/11 & 8847.43 & 0.110 & 0.017& 11&  \\
8854.11 & He I & 5/11? & 8853.59 & 0.047 & 0.007& 13&  \\
8862.26 & H I & P11 & 8862.26 & 9.615 & 1.449& 10& \\
8891.91 & [Fe II]& 13F & 8891.51 & 0.079 & 0.012& 12& \\
8894.21 & O II & & 8893.46 & 0.033 & 0.005& 16& g \\
8914.77 & He I & 2/7 & 8914.22 & 0.152 & 0.023& 10&  \\
8930.97 & He I & 10/11 & 8930.11 & 0.047 & 0.007& 14&  \\
8996.99 & He I & 6/10 & 8996.45 & 0.485 & 0.070& 10&  \\
8999.40 & He I & 7/10 & 8999.16 & 0.142 & 0.021& 11&  \\
9014.91 & H I & P10 & 9014.39 & 13.007 & 1.571& 11& d\\
9063.29 & He I & 4/8 & 9062.71 & 0.302 & 0.036& 11&  \\
9068.90 & [S III] & 1F & 9068.39 &272.159 & 32.644& 11& \\
9085.13 & He I &10/10 & 9084.81 & 0.083 & 0.010& 12&  \\
9095.10 & Ar II & & 9094.68 & 0.080 & 0.010& 12& g \\
9123.60 &[Cl II]& & 9123.17 & 0.212 & 0.025& 11& \\
9204.17 & O II & & 9203.65 & 0.108 & 0.013& 12& \\
9210.28 & He I & 6/9 & 9209.79 & 0.676 & 0.080& 11&  \\
9213.20 & He I & 7/9 & 9212.59 & 0.187 & 0.022& 11&  \\
9229.01 & H I & P9 & 9228.45 & 18.038 & 2.120& 11& d \\
9303.42 & He I & 10/9 & 9302.86 & 0.187 & 0.022& 11& d \\
9463.57 & He I & 1/5 & 9463.10 & 1.096 & 0.125& 11&  \\
9516.57 & He I & 4/7 & 9515.93 & 0.585 & 0.066& 11& d \\
9526.16 & He I & 6/8 & 9526.09 & 0.910 & 0.103& 11&  \\
9530.60 & [S III] & 1F & 9530.42 &709.953 & 80.424& 11& d \\
 & ? & & 9537.38  & 0.239 & 0.027& 11 & d\\
9545.97 & H I & P8 & 9545.47 & 17.376 & 1.965& 11& d \\
9603.44 & He I & 2/6 & 9602.81 & 0.256 & 0.029& 11&  \\
 & ? & & 9822.48  & 0.029 & 0.003& 18&  \\
9824.13 & [C I]& & 9823.77  & 0.287 & 0.031& 11 & \\
9850.26 & [C I]& & 9849.93  & 0.861 & 0.094& 11&  \\
9903.46& C II & 17.02& 9902.88 & 0.738 & 0.080& 11& \\
10027.70 & He I & 6/7 & 10027.12 & 2.065 & 0.221& 11&  \\
10031.20 & He I & 7/7 & 10030.55 & 0.735 & 0.079& 11&  \\
10049.37 & H I & P7 & 10048.79 & 54.885 & 5.853& 11& \\
10138.42 & He I & 10/7 & 10137.81 & 0.283 & 0.030& 12&  \\
10286.73 & [S II] & 3F & 10286.08 & 1.084 & 0.113& 11& \\
10310.70 & He I & 4/6 & 10310.37 & 2.796 & 0.290& 11& c \\
10320.49 & [S II] & 3F & 10319.93 & 3.016 & 0.312& 11& \\
10336.41 & [S II] & 3F & 10335.85 & 2.528 & 0.261& 11& \\
10340.83 & O I & & 10340.39 & 0.531 & 0.055& 11& \\
10370.50 & [S II] & 3F & 10369.95 & 1.114 & 0.115& 11& \\
\enddata
\tablenotetext{a}{Where $F$ is the observed flux in units of 
$100.00=1.056 \times 10^{-12}$ ergs cm$^{-2}$ s$^{-1}$.}
\tablenotetext{b}{Where $I$ is the reddened corrected flux, 
with C(H$\beta$)=1.40 dex, in units of 
$100.00=2.653 \times 10^{-11}$ ergs cm$^{-2}$ s$^{-1}$.}
\tablenotetext{c}{Affected by telluric emission lines.}
\tablenotetext{d}{Affected by atmospheric absorption bands.} 
\tablenotetext{e}{Affected by internal reflections or charge transfer in the CCD.}
\tablenotetext{f}{Blend with an unknown line.} 
\tablenotetext{g}{Dubious identification.} 
\end{deluxetable}

\clearpage
 
\begin{deluxetable}{l@{\hspace{10pt}}l@{\hspace{10pt}}l@{}} 
\tabletypesize{\scriptsize}
\tablecaption{Physical Conditions.
\label{plasma}}
\tablewidth{0pt}
\tablehead{
\colhead{Diagnostic} & 
\colhead{Line}  & }  
\startdata
$N_{\rm e}$ (cm$^{-3}$)& [O\thinspace II] ($\lambda$3726)/($\lambda$3729)& 950 $\pm$ 100 \\
& [O\thinspace II] ($\lambda$3726+$\lambda$3729)/($\lambda$7320+$\lambda$7330)& 2300 $\pm$ 
200$^{\rm a}$ \\
& [S\thinspace II] ($\lambda$6716)/($\lambda$6731)& 1300 $^{+500}_{-300}$ \\ 
& [Fe III] & 3200$\pm$ 400 \\
& [Cl\thinspace III] ($\lambda$5518)/($\lambda$5538)& 3500$^{+900}_{-700}$ \\ 
& [Ar\thinspace IV] ($\lambda$4711)/($\lambda$4740) & 4500$^{+2600}_{-1800}$ \\ 
& Adopted value & 2800$\pm$400 \\
& & \\
$T_{\rm e}$ (K)& [N\thinspace II] ($\lambda$6548+$\lambda$6583)/($\lambda$5755)& 8500 $\pm$ 
200$^{\rm a}$ \\
& [S\thinspace II] ($\lambda$6716+$\lambda$6731)/($\lambda$4069+$\lambda$4076)&  8400 
$^{+350}_{-600}$ \\
& [O\thinspace III] ($\lambda$4959+$\lambda$5007)/($\lambda$4363) & 8500 $\pm$ 50 \\
& [Ar\thinspace III] ($\lambda$7136+$\lambda$7751)/($\lambda$5192)& 8600 $^{+450}_{-350}$ \\
& [S\thinspace III] ($\lambda$9069+$\lambda$9532)/($\lambda$6312) & 9300 $^{+500}_{-400}$ \\
& Adopted value & 8500$\pm$150 \\
& He\thinspace II & 6800$\pm$400 \\
& Balmer decrement & 6650$\pm$750\\ 
& Paschen decrement & 6700$\pm$900 \\ 
\enddata
\tablenotetext{a}{Recombination contribution to the intensity of the auroral lines subtracted (see text).} 
\end{deluxetable}

\begin{deluxetable}{c@{\hspace{10pt}}c@{\hspace{10pt}}}
\tabletypesize{\scriptsize}
\tablecaption{$t^2$ parameter.
\label{t2}}
\tablewidth{0pt}
\tablehead{
\colhead{Method} &
\colhead{$t^2$}} 
\startdata
O$^{\rm ++}$ (R/C)& 0.038$\pm$0.006 \\
Ne$^{\rm ++}$ (R/C)& 0.036$^{+0.014}_{-0.024}$ \\
FL--Pac & 0.038$^{+0.013}_{-0.019}$\\ 
FL--Bac & 0.036$\pm$0.017 \\
Adopted & 0.038$\pm$0.009\\ 
\enddata
\end{deluxetable}

\begin{deluxetable}{ccc}
\tabletypesize{\scriptsize}
\tablecaption{He$^{+}$/H$^{+}$ ratios from singlet lines.
\label{abhe}}
\tablewidth{0pt}
\tablehead{
\colhead{$\lambda_0$(\AA)} &
\colhead{He$^{+}$/H$^{+}$} &
\colhead{12+log(He$^{+}$/H$^{+}$)} } 
\startdata
5015.47& 0.0827$\pm$0.0013& 10.92 \\
3964.50& 0.0919$\pm$0.0031& 10.96 \\
3613.42& 0.0935$\pm$0.0081& 10.97 \\
6677.76& 0.0804$\pm$0.0045& 10.91 \\
4921.69& 0.0836$\pm$0.0014& 10.92 \\
4387.67& 0.0877$\pm$0.0023& 10.94 \\
4009.01& 0.0769$\pm$0.0050& 10.89 \\
4143.52& 0.0830$\pm$0.0037& 10.92 \\
4437.29& 0.1160$\pm$0.0114& 11.06 \\
7280.92& 0.1049$\pm$0.0089& 11.02 \\ \hline
& & \\
Adopted& 0.0866$\pm$0.0080&10.94$\pm$0.04  \\ 
\enddata
\end{deluxetable}

\begin{deluxetable}{l@{}c@{\hspace{10pt}}c@{\hspace{10pt}}} 
\tabletypesize{\scriptsize}
\tablecaption{ Ionic abundances from collisional excited lines$^{\rm a}$.
\label{celabun}}
\tablewidth{0pt}
\tablehead{
\colhead{X$^{\rm m}$} &
\colhead{$t^2$=0.000} &
\colhead{$t^2$=0.038$\pm$0.009}} 
\startdata
N$^{+}$&7.09$\pm$0.06 & 7.25$\pm$0.06 \\
O$^{+}$&8.15$\pm$0.07& 8.32$\pm$0.07 \\
O$^{++}$& 8.35$\pm$0.03 & 8.63$\pm$0.08 \\
Ne$^{++}$& 7.61$\pm$0.09 & 7.91$\pm$0.10 \\
S$^{+}$&5.75$\pm$0.08 & 5.91$\pm$0.08 \\
S$^{++}$& 6.99$\pm$0.10 & 7.30$\pm$0.10 \\
Cl$^{+}$& 4.13$\pm$0.08& 4.26$\pm$0.08 \\
Cl$^{++}$& 4.95$\pm$0.06& 5.21$^{+0.09}_{-0.07}$ \\
Cl$^{3+}$& 3.21$\pm$0.07& 3.42$\pm$0.08 \\
Ar$^{++}$& 6.34$\pm$0.05 & 6.57$\pm$0.08 \\
Ar$^{3+}$& 4.20$\pm$0.07& 4.48$\pm$0.09 \\
Fe$^{++}$& 5.57$\pm$0.05& 5.85$\pm$0.09\\
Fe$^{3+}$& 5.71$^{+0.17}_{-0.29}$& 5.95$^{+0.12}_{-0.16}$\\ 
\enddata
\tablenotetext{a}{In units of 12+log(X$^{\it m}$/H$^{\it +}$).}
\end{deluxetable}

\begin{deluxetable}{c@{\hspace{10pt}}c@{\hspace{10pt}}c@{\hspace{10pt}}c@{\hspace{10pt}}c@{\hspace{10pt}}} 
\tabletypesize{\scriptsize}
\tablecaption{ C$^{\it ++}$/H$^{\it +}$ ratio from C II lines.
\label{cii}}
\tablewidth{0pt}
\tablehead{
\colhead{Mult} &
\colhead{$\lambda$ (\AA)} &
\colhead{$I$($\lambda$)/$I$(H$\beta$)} &
\multicolumn{2}{c}{C$^{++}$/H$^{+}$ ($\times$ 10$^{-5}$)} \\
& & \colhead{[$I$(H$\beta$)=100]} &
\colhead{A} & \colhead{B}} 
\startdata
2& 6578.05$^{\rm a}$& 0.224& 269& 46 \\
3& 7231.12& 0.093& 2408& 34 \\
& 7236.19& 0.148& 2176& 31 \\
& Average& & 2266& 32 \\
& m$_{\rm cf}$=1.09& & & \\
& Sum& 0.257& 2260& 32 \\
6& 4267.26& 0.295& 27& {\bf 27} \\ 
16.04& 6151.43& 0.012& {\bf 27}& 26 \\ 
17.02& 9903.46& 0.080& {\bf 29}& -- \\ 
17.04& 6462.00& 0.032& {\bf 28}& -- \\ 
17.06& 5342.38& 0.012:& {\bf 20}& -- \\ \hline
& & & & \\ 
& Adopted& &\multicolumn{2}{c}{\bf 28 $\pm$ 4 } \\ 
\enddata
\tablenotetext{a}{Affected by a telluric emission line.} 
\end{deluxetable}

\begin{deluxetable}{c@{}c@{}c@{}c@{}c@{}c@{}c@{}}
\tabletypesize{\scriptsize}
\tablecaption{Nitrogen abundances from permitted lines$^{\rm a}$. 
\label{nitro}}
\tablewidth{0pt}
\tablehead{
\colhead{ion} &
\colhead{Mult} &
\colhead{$\lambda$ (\AA)} &
\colhead{$I$($\lambda$)/$I$(H$\beta$)} &
\multicolumn{2}{c}{X$^{+i}$/H$^{+}$ ($\times$ 10$^{-5}$)} &
\colhead{notes} \\
& & & \colhead{[$I$(H$\beta$)=100]} & \colhead{A}& \colhead{B}  } 
\startdata
N$^{\rm +}$ & 1& 8680.28& 0.042& 117& 114& \\
& & 8683.40 & 0.028& 151 & 147& \\
& & 8686.15 & 0.029& 393 & 381& \\
& & 8703.25 & 0.019& 241 & 234& \\
& & 8711.70 & 0.023& 237 & 230& \\
& & 8718.84 & 0.012& 152 & 147& \\
& & Average & &212 & 205& \\
& & m$_{\rm cf}$=1.02 & & & & \\
& & Sum & & 175 & 170 & \\
& 2 &   8184.85 & 0.018& 281 & 239&   \\
& & 8188.01 & 0.036&570 & 484& \\
& & 8210.72 & 0.010&389 & 331& \\
& & 8216.28 & 0.044&266 & 226& \\
& & Average & &381 & 323& \\
& & m$_{\rm cf}$=1.48 & & & & \\
& & Sum & & 339 & 288&  \\
& 3 & 7423.64 & 0.014& 1360  & 444&  \\
& & 7442.30 & 0.028& 1348  & 440& \\
& & 7468.31 & 0.042& 1353  & 441& \\
& & Average & &1353  & 441& \\
& & m$_{\rm cf}$=1.00 & & & & \\
& & Sum & & 1353  & 441& \\ \hline

N$^{\rm ++}$ & 3 & 5666.64 & 0.021& 12 & 10& KS02 \\
& & 5676.02 & 0.014&18 & 15& \\
& & 5679.56 & 0.039&12 & 10& \\
& & 5686.21 & 0.005:&8 & 7& \\
& & 5710.76 & 0.005:&8 & 7& \\
& & Average & &13 & 11& \\
& & m$_{\rm cf}$=1.07& & & & \\
& & Sum & & 12 & {\bf 10} & \\
& 5 & 4601.48 & 0.013& 107 & 18& KS02 \\
& & 4613.87 & 0.010&188 & 32& \\
& & 4621.39 & 0.023&272 & 46& \\
& & 4630.54 & 0.055&144 & 24& \\
& & 4643.06 & 0.027&212 & 36& \\
& & Average & &181 & 30& \\
& & m$_{\rm cf}$=1.12& & & & \\
& & Sum & & 164 & 28& \\
& 20 & 4803.29 & 0.019& 1195  & 23& KS02\\
& & 4779.71& 0.009& 1615& 23& \\
& & 4788.13& 0.017& 1957& 38& \\
& & Average & & 1573& 31& \\
& & m$_{\rm cf}$=1.27& & & & \\
& & Sum& & 1498& 29& \\
& 28 & 5927.82 & 0.009 & 2629 & 31& KS02\\
& & 5931.79& 0.020& 2568& 30& \\
& & 5940.24& 0.006:& 2240& 26& \\
& & 5941.68& 0.015& 1017& 12& \\
& & Average & & 2059& 24& \\
& & m$_{\rm cf}$=1.10& & & & \\
& & Sum& & 1703& 20& \\
& 48 & 4239.4 & 0.034& 8 & {\bf 8}& EV90 \\ \hline 
& & & & & & \\
& Adopted& & & \multicolumn{2}{c}{\bf 10 $\pm$ 1}& \\ 
\enddata
\tablenotetext{a}{Only lines with intensity uncertainties lower than 40\% have been considered.} 
\tablenotetext{b}{Recombination coefficients from: KS02 = \citet{kiss02}, EV90 = \citet{esc90}.} 
\end{deluxetable}

\begin{deluxetable}{c@{\hspace{10pt}}c@{\hspace{10pt}}c@{\hspace{10pt}}c@{\hspace
{10pt}}c@{\hspace{10pt}}}
\tabletypesize{\scriptsize}
\tablecaption{O$^+$/H$^+$ ratio from \ion{O}{1} lines.
\label{oi}}
\tablewidth{0pt}
\tablehead{
\colhead{Mult} &
\colhead{$\lambda$ (\AA)} &
\colhead{$I$($\lambda$)/$I$(H$\beta$)} &
\multicolumn{2}{c}{O$^{+}$/H$^{+}$ $(\times 10^{-5})^{\rm a}$}  \\
& & \colhead{[$I$(H$\beta$)=100]} & \colhead{A}& \colhead{B}}
\startdata
1& 7771.94& 0.011& {\bf 11/15}& -- \\
& 7774.17$^{\rm b}$& 0.056& 80/103 & -- \\
& & & & \\
4& 8446.48$^{\rm b}$& 0.638& 2425/3641& 548/730 \\ \hline
& & & & \\ 
& Adopted& &\multicolumn{2}{c}{\bf 13 $\pm$ 3 } \\ 
\tablenotetext{a}{Recombination coefficients from Pequignot et al. (1991) / Escalante
\& Victor (1992).}
\tablenotetext{b}{Affected by telluric emission lines.} 
\enddata
\end{deluxetable}

\begin{deluxetable}{c@{\hspace{10pt}}c@{\hspace{10pt}}c@{\hspace{10pt}}c@{\hspace{10pt}}c@{\hspace
{10pt}}c@{\hspace{10pt}}}
\tabletypesize{\scriptsize}
\tablecaption{O$^{\it ++}$/H$^{\it +}$ ratio from O II lines$^{\rm a}$. 
\label{oii}}
\tablewidth{0pt}
\tablehead{
\colhead{Mult} &
\colhead{$\lambda$ (\AA)} &
\colhead{$I$($\lambda$)/$I$(H$\beta$)} &
\multicolumn{3}{c}{O$^{++}$/H$^{+}$ ($\times$ 10$^{-5}$)} \\
& & \colhead{[$I$(H$\beta$)=100]} & \colhead{A}& \colhead{B}& \colhead{C} } 
\startdata
1& 4638.85& 0.074& 73& 70&- -- \\
& 4641.81& 0.132& 47& 45& -- \\ 
& 4649.14& 0.145& 29& 28& -- \\
& 4650.84& 0.069& 70& 68& -- \\
& 4661.64& 0.090& 72& 69& -- \\
& 4673.73& 0.013 & 84& 81& -- \\
& 4676.23& 0.040& 43& 41& -- \\
& 4696.36& 0.007: & 79& 76& -- \\
& Average& & 53& 51& -- \\
& m$_{\rm cf}$=1.00& & & & \\
& Sum& & 46& {\bf 45}& -- \\
2& 4317.14& 0.026& 48& 35&- -- \\
& 4319.63& 0.019& 37& 26& -- \\
& 4345.56$^{\rm b}$& 0.057& 103 & 73 & -- \\
& 4349.43& 0.067& 48& 34& -- \\
& 4366.89& 0.040& 63& 45& -- \\
& Average& & 51& 36& -- \\
& m$_{\rm cf}$=1.28& & & & \\
& Sum& & 49& {\bf 35}& -- \\
4& 6721.39& 0.004:& 60& --& 47 \\
& m$_{\rm cf}$=1.50& & & & \\
& Sum& & {\bf 60}& -- & 47 \\
5& 4414.90& 0.029& 56& --& 9 \\
& 4416.97& 0.031& 108& --& 17 \\
& Average& & 82& --& 13 \\
& m$_{\rm cf}$=1.07& & & & \\
& Sum& & {\bf 74}& --& 12 \\
$\left.\matrix{10^{\rm c} \cr  }\right.$ &
$\left.\matrix{4069.62\cr 4069.89}\right.$ & 
0.139 & 54/52& --& -- \\
& 4072.15& 0.085& 35/34& --& -- \\
& 4075.86& 0.108& 30/30& --& -- \\
& 4085.11& 0.024& 47/53& --& -- \\
-& Average& & 42/41& -- & -- \\
& m$_{\rm cf}$=1.10/1.25& & & & \\
& Sum& & 39/{\bf 39}& --& -- \\
15$^{\rm d}$& 4590.97& 0.029& 164& --& -- \\
$\left.\matrix{  \cr  }\right.$ &
$\left.\matrix{4595.95\cr 4596.18}\right.$ & 
0.010& 190& --& -- \\
& Average& &176& --& -- \\
& m$_{\rm cf}$=1.00& & & & \\
& Sum& & 176 & --& -- \\
19$^{\rm c}$& 4129.32& 0.009:& 4647/2219 & 175/135& 175/126 \\
& 4132.80& 0.035& 1560/1181 & 59/61& 59/57 \\ 
& 4153.30& 0.043& 1419/1240 & 54/54& 54/50 \\
& 4156.54$^{\rm f}$& 0.028& 6744/4005 & 255/219& 255/205 \\
& Average& & 1482/1214 & 56/57& 56/54 \\
& m$_{\rm cf}$=1.37/1.47& & & & \\
& Sum& & 1481/1215 & 56/{\bf 57} & 56/53 \\
20$^{\rm c}$& 4110.78& 0.015& 195/455 & 190/60& 190/57 \\
& 4119.22& 0.023& 13/26 & 12/25 & 12/13 \\ 
& Average& &  85/196 & 83/39& 83/31 \\
& m$_{\rm cf}$=2.42/1.89& & & & \\
& Sum& & 20/41 & 19/{\bf 33} & 19/19 \\
25$^{\rm c}$& 4699.21& 0.007:& 100/105& 100/91& 5/9 \\
& 4705.35& 0.008:& 77/69& 77/67& 4/4 \\ 
& Average& & 88/86& 88/79& 4/7 \\
& m$_{\rm cf}$=1.03/1.04& & & & \\
& Sum& & 87/82 & 87/77 & 4/5 \\
3d--4f$^{\rm e}$& 4083.90& 0.017& --& 52& -- \\
& 4087.15& 0.026& --& 81& -- \\
& 4089.29& 0.044& --& 37& -- \\
& 4275.55& 0.019& --& 30& -- \\
& 4285.69& 0.015& --& 66& -- \\
& 4332.71& 0.013:& --& 116& -- \\
& 4491.23$^{\rm b}$& 0.022& --& 137& -- \\
& 4602.13& 0.006:& --& 28& -- \\
& 4609.44& 0.024& --& 48& -- \\
& Average& & --& {\bf 50}& -- \\ \hline
& & & & & \\ 
& Adopted& &\multicolumn{3}{c}{\bf 42 $\pm$ 5 } \\ 
\tablenotetext{a}{Only lines with intensity uncertainties lower than 40 \% have been considered.} 
\tablenotetext{b}{Blend.} 
\tablenotetext{c}{Values for LS coupling (left) and intermediate coupling (right).} 
\tablenotetext{d}{Dielectronic recombination coefficients (Nussbaumer \& Storey, 1984).} 
\tablenotetext{e}{Values for Intermediate coupling.} 
\tablenotetext{f}{Probably blended (see text).} 
\enddata
\end{deluxetable}

\begin{deluxetable}{c@{\hspace{10pt}}c@{\hspace{10pt}}c@{\hspace{10pt}}c@{\hspace{10pt}}c@{\hspace{10pt}}} 
\tabletypesize{\scriptsize}
\tablecaption{ Ne$^{\it ++}$/H$^{\it +}$ ratio from Ne II lines.
\label{neii}}
\tablewidth{0pt}
\tablehead{
\colhead{Mult} &
\colhead{$\lambda$ (\AA)} &
\colhead{$I$($\lambda$)/$I$(H$\beta$)} &
\multicolumn{2}{c}{Ne$^{++}$/H$^{+}$ ($\times$ 10$^{-6}$)} \\
& & \colhead{[$I$(H$\beta$)=100]} &
\colhead{A} & \colhead{B}} 
\startdata
55e& 4391.99& 0.018:& 55& -- \\
& 4409.30& 0.024:& 111& -- \\
& Average& & 87& -- \\
& m$_{\rm cf}$=1.00& & & \\
& Sum& 0.042& {\bf 77}& -- \\ \hline
& & & & \\ 
& Adopted& &\multicolumn{2}{c}{\bf 77 $\pm$ 25$^{\rm a}$ } \\ 
\tablenotetext{a}{Assuming a 50 \% error in line intensities.} 
\enddata
\end{deluxetable}

\begin{deluxetable}{l@{}c@{\hspace{10pt}}c@{\hspace{10pt}}} 
\tabletypesize{\scriptsize}
\tablecaption{ Total gaseous abundances$^{\rm a}$.
\label{totabun}}
\tablewidth{0pt}
\tablehead{
\colhead{Element} &
\colhead{$t^2$=0.000} &
\colhead{$t^2$=0.038}} 
\startdata
He&10.97$\pm$ 0.03& 10.97$\pm$ 0.04 \\
C$^{\rm b}$&8.47$\pm$ 0.06 & 8.46$\pm$ 0.06 \\
N&7.63$\pm$0.06 & 7.87$\pm$0.07\\
N$^{\rm c}$& ---- & 8.07$\pm$0.08\\
O& 8.56$\pm$0.03 & 8.80$\pm$0.07 \\
O$^{\rm b}$& 8.74$\pm$0.06 & 8.74$\pm$0.06 \\
Ne& 7.82$\pm$0.10 & 8.09$\pm$0.12 \\
Ne$^{\rm b}$& 8.09$\pm$0.10 & 8.07$\pm$0.11 \\
S& 7.05$\pm$0.10 & 7.36$\pm$0.10 \\
Cl& 5.02$\pm$0.06& 5.26$\pm$0.08 \\
Ar& 6.38$\pm$0.06& 6.61$\pm$0.08 \\
Fe$^{\rm d}$& 5.96$^{+0.09}_{-0.11}$ & 6.30$\pm$0.14 \\ 
Fe$^{\rm e}$& 5.95$^{+0.11}_{-0.15}$ & 6.20$\pm$0.12 \\ 
\enddata
\tablenotetext{a}{In units of 12+log(X/H).} 
\tablenotetext{b}{From recombination lines (RLs).}
\tablenotetext{c}{N$^+$/H$^+$ from collisional excited lines (CELs) and
N$^{++}$/H$^+$ from permitted lines.} 
\tablenotetext{d}{Assuming $ICF$(Fe).} 
\tablenotetext{e}{Fe/H= Fe$^{\rm ++}$/H$^{\rm +}$+Fe$^{\rm 3+}$/H$^{\rm +}$.} 
\end{deluxetable}

\begin{deluxetable}{l@{}c@{\hspace{10pt}}c@{\hspace{10pt}}c@{\hspace{10pt}}c@
{\hspace{10pt}}c@{\hspace{10pt}}} 
\tabletypesize{\scriptsize}
\tablecaption{ Comparison of NGC 3576 gaseous abundance determinations$^{a}$.
\label{comparison}}
\tablewidth{0pt}
\tablehead{
\colhead{Element} &
\colhead{This work ($t^2$=0.00)} &
\colhead{(1)} &
\colhead{(2)} &
\colhead{(3)} & 
\colhead{(4)}} 
\startdata
& $\alpha$= 11:12:0.9&  $\alpha$=11:12:0.5& $\alpha$=11:14:50.1.4& 
$\alpha$=11:12:48.1& $\alpha$=11:11:46.6 \\
& $\delta$=-61:18:19.1& $\delta$=-61:18:24& $\delta$=-61:37:35.3& 
$\delta$=-61:33:12.3& $\delta$=-61:18:43 \\ \hline            
He$^{+}$ & 10.97  & 10.97 & 10.94& 10.98& ----- \\
C$^{++}$$^{\rm b}$ & 8.47  & 8.46 & ----- & ----- & -----     \\
N$^{+}$  & 7.09 & 7.07 &  7.45  & 7.20 & ----- \\
N$^{++}$  & 8.00$^{\rm c}$  & 8.43$^{\rm c}$ &  ----- & ----- & 7.68 \\
N   & 7.63$^{\rm d}$  & 7.55 $^{\rm d}$&  7.58  & 7.56  &7.85 \\
O$^{+}$  & 8.15 & 8.04 &  8.55  & 8.18  & ----- \\
O$^{+}$$^{\rm b}$  & 8.11       & ----- & -----  & -----  & ----- \\
O$^{++}$ & 8.35 & 8.34 &  8.08  & 8.30  & 8.43 \\
O$^{++}$$^{\rm b}$ & 8.62       & 8.57 &  ----  & ----  & ---- \\
O$^{\rm e}$   & 8.56  & 8.52 &  8.67  & 8.55  & ----- \\
Ne$^{++}$& 7.61 & 7.54 &   7.16  & 7.35  & 7.70 \\
Ne  & 7.82  & 7.72 &  7.75  & 7.40  & 8.00 \\
S$^{+}$  & 5.67 & 5.79 &  6.16  & 5.84  & ----- \\
S$^{++}$ & 6.99 & ----- &  6.82  & 6.80  & 6.86     \\
S   & 7.05  & $\ge$5.82 &   7.03  & 7.49  &7.04 \\
Cl$^{+}$$^{\rm f}$ & 4.13 & ----- & ----- &----- &----- \\
Cl$^{++}$& 4.95 & 5.00 &  ----- & ----- &----- \\
Cl$^{3+}$& 3.21 & ----- &  ----- & ----- & ----- \\
Cl  & 5.02 & 5.18  &  ----- & ----- & ----- \\
Ar$^{++}$& 6.34 & 6.23 &  5.98  & 6.21  & ----- \\
Ar$^{3+}$& 4.20 & 4.33 &   ----- & ----- & ----- \\
Ar  & 6.38 & 6.41 & ----- & ----- & ----- \\ 
\enddata
\tablenotetext{a}{REFERENCES.- (1) Tsamis et al. (2002); (2) (3) Girardi et al. (1997); (4) 
Simpson et al. (1995).} 
\tablenotetext{b}{Abundances from RLs.} 
\tablenotetext{c}{Abundances from permitted lines probably affected by fluorescence.} 
\tablenotetext{d}{\emph{ICF} assumed.} 
\tablenotetext{e}{Only from CELs.} 
\tablenotetext{f}{Atomic data not reliable (see text).} 
\end{deluxetable}

\begin{deluxetable}{l@{}c@{\hspace{10pt}}c@{\hspace{10pt}}c@{\hspace{10pt}}} 
\tabletypesize{\scriptsize}
\tablecaption{NGC 3576 and Solar abundances$^{\rm a}$.
\label{solar}}
\tablewidth{0pt}
\tablehead{
\colhead{Element} &
\colhead{NGC 3576} &
\colhead{Sun$^{\rm c}$} &
\colhead{NGC 3576--Sun}} 
\startdata
He          &10.97$\pm$0.04  &10.98$\pm$0.02 & --0.01 \\
C$^{\rm b}$ & 8.56$\pm$0.06  & 8.41$\pm$0.05 & +0.15 \\
N           & 7.86$\pm$0.07  & 7.80$\pm$0.05 & +0.06 \\
O$^{\rm b}$ & 8.82$\pm$0.06  & 8.66$\pm$0.05 & +0.16 \\
Ne          & 8.08$\pm$0.12  & 7.84$\pm$0.06 & +0.24 \\
S           & 7.36$\pm$0.10  & 7.20$\pm$0.08 & +0.16 \\
Cl          & 5.26$\pm$0.08  & 5.28$\pm$0.08 & --0.02 \\
Ar          & 6.61$\pm$0.08  & 6.18$\pm$0.08 & +0.43 \\
Fe          & 7.66$\pm$0.20  & 7.50$\pm$0.05 & +0.16 \\ 
\enddata
\tablenotetext{a}{In units of 12+log(X/H).} 
\tablenotetext{b}{Values derived from RLs.} 
\tablenotetext{c}{\citet{chr98,gre98,asp03,asp04}.}
\end{deluxetable}

\begin{thebibliography}{}

\bibitem[Akerman et al.(2004)]{ake04}
Akerman, C. J., Carigi, L., Nissen, P. E., Pettini, M., \&
Asplund, M. 2004,
\aap, 414, 931

\bibitem[Asplund (2003)]{asp03} 
Asplund, M. 2003,
in:  ASP Conference Series 304, CNO in the Universe, 
ed. C. Charbonnel, D, Schaerer, \& G. Meynet, 275

\bibitem[Asplund et al.(2004)]{asp04} 
Asplund, M., Grevesse, N., Sauval, A. J, Allende-Prieto, C., \& 
Kiselman, D. 2004,
\aap, in press, (astro-ph/0312290)

\bibitem[Baldwin et al.(2000)]{bald00} 
Baldwin, J.A., Verner, E.M., Verner, D.A., Ferland, G.J., 
Martin, P.G., Korista, K.T., \& Rubin, R.H. 2000, 
\apjs, 129, 229

\bibitem[Bautista \& Pradhan(1996)]{bau96}
Bautista, M.A., \& Pradhan, A.K. 1996,
A\&AS, 115, 551

\bibitem[Benjamin, Skillman, \& Smits(1999)]{ben99} 
Benjamin, R.A., Skillman, E.D., \& Smits, D.P. 1999, 
\apj, 514, 307

\bibitem[Boreiko \& Betz(1997)]{bor97}
Boreiko, R.T., \& Betz, A.L. 1997, 
\apjs, 111, 409

\bibitem[Brown \& Mathews(1970)]{brown70} 
Brown, R.L., \& Mathews, W.G. 1970, 
\apj, 160, 939

\bibitem[Christensen-Dalsgaard(1998)]{chr98} 
Christensen-Dalsgaard, J. 1998, 
Space Sci. Rev., 85, 19

\bibitem[Carigi(2003)]{car03}
Carigi, L. 2003,
\mnras, 339, 825

\bibitem[Davey, Storey, \& Kisielius(2000)]{dav00} 
Davey, A.R., Storey, P.J., \& Kisielius, R. 2000, 
\aap, 142, 85

\bibitem[Deharveng et al.(2000)]{dehar00} 
Deharveng, L., Pe\~na, M., Caplan, J., \& Costero, R. 2000, 
\mnras, 311, 329

\bibitem[De Robertis, Dufour, \& Hunt(1987)]{derob87} 
De Robertis, M.M., Dufour, R.J., \& Hunt, R.W. 1987, 
\jrasc, 81, 195

\bibitem[D'Odorico et al.(2000)]{dodo00} 
D'Odorico, S., Cristiani, S., Dekker, H., Hill, V., Kaufer, A., 
Kim, T., \& Primas, F. 2000, 
Proc. SPIE, 4005, 121

\bibitem[Escalante \& Victor(1990)]{esc90} 
Escalante, V., \& Victor, G.A. 1990, 
\apjs, 73, 513

\bibitem[Escalante \& Victor(1992)]{esc92} 
Escalante, V., \& Victor, G.A. 1992, 
\planss, 40, 1705

\bibitem[Esteban(2002)]{este02} 
Esteban, C. 2002, 
Rev. Mexicana. Astron. Astrof\'{\i}s. Ser. Conf., 12, 56 

\bibitem[Esteban et al.(2004)]{este04} 
Esteban, C., Peimbert, M., Garc\'{\i}a-Rojas, J., Peimbert, A., Ruiz, M.T., \& Rodriguez, M. 2004, 
in preparation

\bibitem[Esteban et al.(1998)]{EPTE} 
Esteban, C., Peimbert, M., Torres-Peimbert, S., \& Escalante, V. 1998, 
\mnras, 295, 401 (EPTE)

\bibitem[Esteban et al.(1999b)]{EPTG} 
Esteban, C., Peimbert, M., Torres-Peimbert, S., \& Garc\'{\i}a-Rojas, J. 1999b, 
Rev. Mexicana. Astron. Astrof\'{\i}s., 35, 65 

\bibitem[Esteban et al.(1999a)]{EPTGR} 
Esteban, C., Peimbert, M., Torres-Peimbert, S., Garc\'{\i}a-Rojas, J., \& Rodr\'{\i}guez, M. 
1999a, 
\apjs, 120, 113 (EPTGR)

\bibitem[Esteban et al.(2002)]{este02b} 
Esteban, C., Peimbert, M., Torres-Peimbert, S., \& Rodr\'{\i}guez, M. 2002, 
\apj, 581, 241

\bibitem[Feklistova et al.(1994)]{fek94} 
Feklistova, T., Golovatyj, V.V., Kholtygin, A.F., \& Sapar, A. 1994, 
Baltic Astronomy, 3, 292

\bibitem[Figu\^eredo et al.(2002)]{figue02} 
Figu\^eredo, E., Blum, R.D., Damineli, A., \& Conti, P.S. 2002, 
\aj, 124, 2739

\bibitem[Froese Fischer \& Rubin(1998)]{fro98} 
Froese Fischer C., Rubin R. H., 1998, 
J. Phys. B: At. Mol. Opt. Phys., 31, 1657

\bibitem[Garnett et al.(1999)]{garn99} 
Garnett, D. R., Shields, G. A., Peimbert, M., Torres-Peimbert, S., Skillman, E. D., 
Dufour, R. J., Terlevich, E., \& Terlevich, R. J., 1999, 
\apj, 513, 168

\bibitem[Garstang(1958)]{gar58} 
Garstang R. H., 1958, 
\mnras, 118, 572

\bibitem[Girardi et al.(1997)]{girardi97} 
Girardi, L., Bica, E., Pastoriza, M.G., \& Winge, C. 1997, 
\apj, 486, 847

\bibitem[Grandi(1975a)]{gra75a} 
Grandi, S.A. 1975a, 
\apj, 166, 465

\bibitem[Grandi(1975b)]{gra75b} 
Grandi, S.A. 1975b, 
\apj, 199, L43

\bibitem[Grandi(1976)]{gra76} 
Grandi, S.A. 1976, 
\apj, 206, 658

\bibitem[Grevesse \& Sauval(1998)]{gre98}
Grevesse, N., \& Sauval, A. J. 1998,
Space Sci. Rev., 85, 161

\bibitem[Herrero(2003)]{herr03} 
Herrero, A., 2003, 
in:  ASP Conference Series 304, CNO in the Universe, 
ed. C. Charbonnel, D, Schaerer, \& G. Meynet, 10

\bibitem[Howard \& Murray(1990)]{how90} 
Howard, I.D., \& Murray, J. 1990, 
SERC Starlink User Note No. 50

\bibitem[Humphreys(1978)]{hum78} 
Humphreys, R.M. 1978, 
\apjs, 38, 309

\bibitem[Kisielius \& Storey(2002)]{kiss02} 
Kisielius, R., \& Storey, P. J. 2002, 
\aap, 387, 1135

\bibitem[Liu(2002)]{liu02} 
Liu, X.-W. 2002, 
Rev. Mexicana. Astron. Astrof\'{\i}s. Ser. Conf., 12, 70 

\bibitem[Liu(2003)]{liu03} 
Liu, X.-W. 2003, 
in IAU Symposium 209, Planetary Nebulae and Their Role in the Universe, 
ed. S. Kwok, M. Dopita, \& R. Sutherland (San Francisco: 
ASP), 339

\bibitem[Liu et al.(2001)]{liu01} 
Liu, X.-W., Luo, S.-G., Barlow, M.J., Danziger, I.J., \& Storey, P.J. 2001, 
\mnras, 327, 141

\bibitem[Liu et al.(1995)]{liu95} 
Liu, X.-W., Storey, P.J., Barlow, M.J., \& Clegg, R.E.S. 1995, 
\mnras, 272, 369

\bibitem[Liu et al.(2000)]{liu00} 
Liu, X.-W., Storey, P.J., Barlow, M.J., Danziger, I.J., Cohen, M., \& Bryce, M. 2000, 
\mnras, 312, 585

\bibitem[Mart\'{\i}n-Hern\'andez et al.(2002)]{martin02} 
Mart\'{\i}n-Hern\'andez, N.L., Peeters, E., Morisset, C., et al. 2002, 
\aap, 381, 606

\bibitem[Mathis \& Rosa(1991)]{mathis91} 
Mathis, J.S., \& Rosa, M.R. 1991, 
\aap, 245, 625

\bibitem[Mendoza(1983)]{men83} 
Mendoza C., 1983, Flower D. R., Reidel D., eds, 
Proc. IAU Symp. 103, Planetary Nebulae, Kluwer, Dordrecht, 143

\bibitem[Moore(1945)]{moo45} 
Moore, C.E. 1945, 
A Multiplet Table of Astrophysical Interest 
(Contributions from the Princenton University Observatory, No. 20. Princenton: The Observatory)

\bibitem[Moore(1993)]{moo93} 
Moore, C.E. 1993,  
Tables of Spectra of Hydrogen, Carbon, Nitrogen and Oxygen Atoms and Ions (Boca Raton: CRC)

\bibitem[Nussbaumer \& Storey(1984)]{nuss84} 
Nussbaumer, H., \& Storey, P.J. 1984, 
A\&AS, 56, 293

\bibitem[O'Dell et al.(2003)]{odell03} 
O'Dell, C.R., Peimbert, M., \& Peimbert, A. 2003, 
\aj, 125, 2590

\bibitem[Osterbrock, Tran, \& Veilleux(1992)]{ost92} 
Osterbrock, D.E., Tran, H.D., \& Veilleux, S. 1992, 
\apj, 389, 305

\bibitem[Peimbert(2003)]{peim03} 
Peimbert, A. 2003, 
\apj, 584, 735

\bibitem[Peimbert, Peimbert, \& Luridiana(2002)]{pepe02} 
Peimbert, A., Peimbert, M., \& Luridiana, V. 2002, 
\apj, 565, 668

\bibitem[Peimbert(1967)]{peim67} 
Peimbert, M. 1967, 
\apj, 150, 825

\bibitem[Peimbert(1971)]{peim71} 
Peimbert, M. 1971, 
Bol. Obs. Tonantzintla y Tacubaya, 6, 29

\bibitem[Peimbert \& Costero(1969)]{peim69} 
Peimbert, M., \& Costero, R. 1969, 
Bol. Obs. Tonantzintla y Tacubaya, 5, 3

\bibitem[Peimbert, Peimbert, \& Ruiz(2000)]{peim00} 
Peimbert, M., Peimbert, A., \& Ruiz, M.T. 2000, 
\apj, 541, 688

\bibitem[Peimbert et al.(2004)]{peim04} 
Peimbert, M., Peimbert, A., Ruiz, M.T., \& Esteban, C. 2004, 
\apjs, 150, 431

\bibitem[Peimbert, Storey, \& Torres-Peimbert(1993)]{peim93a} 
Peimbert, M., Storey, P.J., \& Torres-Peimbert, S. 1993, 
\apj, 414, 626

\bibitem[Peimbert \& Torres-Peimbert(1977)]{peim77} 
Peimbert, M., \& Torres-Peimbert, S. 1977, 
\mnras, 179, 217

\bibitem[Peimbert, Torres-Peimbert, \& Ruiz(1992)]{peim92} 
Peimbert, M., Torres-Peimbert, S., \& Ruiz, M.T. 1992, 
Rev. Mexicana. Astron. Astrof\'{\i}s., 24, 155

\bibitem[P\'equignot \& Baluteau(1988)]{peq88} 
P\'equignot, D., \& Baluteau, J.-P. 1988, 
\aap, 206, 298

\bibitem[P\'equignot, Petitjean, \& Boisson(1991)]{peq91} 
P\'equignot, D., Petitjean, P., \& Boisson, C. 1991, 
\aap, 251, 680

\bibitem[Quinet(1996)]{qui96} 
Quinet P., 1996, 
A\&AS, 116, 573

\bibitem[Rodgers, Campbell, \& Whiteoak(1960)]{rodg60} 
Rodgers, A.W., Campbell, C.T., \& Whiteoak, J.B. 1960, 
\mnras, 121, 103 

\bibitem[Rodr\'{\i}guez(1996)]{rodri96} 
Rodr\'{\i}guez, M. 1996, 
\aap, 313, L5

\bibitem[Rodr\'{\i}guez(1999)]{rodri99} 
Rodr\'{\i}guez, M. 1999, 
\aap, 348, 222

\bibitem[Rodr\'{\i}guez(2002)]{rodri02} 
Rodr\'{\i}guez, M. 2002, 
\aap, 389, 556

\bibitem[Rodr\'{\i}guez(2003)]{rodri03} 
Rodr\'{\i}guez, M. 2003, 
\apj, 590, 296

\bibitem[Rodr\'{\i}guez \& Rubin(2004)]{rodri04} 
Rodr\'{\i}guez, M., \& Rubin, R.H. 2004, 
in Recycling Intergallactic and Interstellar Matter, 
IAU Symposium No 217, in press (astro-ph/0312246)

\bibitem[Rolleston et al. (2000)]{roll00}
Rolleston, W.R.J., Smartt, S.J., Dufton, P.L., \& Ryans, R.S.I. 2000,
\aap, 363, 537

\bibitem[Rubin et al.(2003)]{rubin03} 
Rubin, R.H., Martin, P.G., Dufour, R.J., Ferland, G.J., Blagrave, K.P.M., 
Liu, X.-W., Nguyen, J.F., \& Baldwin, J.A. 2003, 
\mnras, 340, 362

\bibitem[Ruiz et al.(2003)]{ruiz03} 
Ruiz, M.T., Peimbert, A., Peimbert, M., \& Esteban, C. 2003, 
\apj, 595, 247

\bibitem[Russeil(2003)]{russ03} 
Russeil, D. 2003, 
\aap, 397, 133

\bibitem[Savage \& Mathis(1979)]{sav79} 
Savage, B.D., \& Mathis, J.S. 1979, 
\araa, 17, 73

\bibitem[Savage \& Sembach(1996)]{sav96} 
Savage, B.D., \& Sembach, K.R. 1996, 
\apj, 457, 211

\bibitem[Sharpee et al.(2003)]{shar03} 
Sharpee, B., Williams, R., Baldwin, J.A., \& van Hoof, P.A.M. 2003, 
\apjs, 149, 157

\bibitem[Shaw \& Dufour(1995)]{shaw95} 
Shaw, R.A., \& Dufour, R. 1995, 
\pasp, 107, 896

\bibitem[Smartt et al.(2001)]{smar01} 
Smartt, S.J., Venn, K.A., Dufton, P.L., Lennon, D.J., Rolleston, W.R.J., \& Keenan, F.P. 2001, 
\aap, 367, 86

\bibitem[Smits(1996)]{smits96} 
Smits, D.P. 1996, 
\mnras, 278, 683

\bibitem[Simpson et al.(1995)]{simpson95} 
Simpson, J.P, Colgan, S.W.J., Rubin, R.H., Erickson, E.F., \& Haas, M.R. 1995, 
\apj, 444, 721

\bibitem[Storey(1994)]{sto94} 
Storey, P. J. 1994, 
\aap, 282, 999

\bibitem[Storey \& Hummer(1995)]{sto95} 
Storey, P. J., \& Hummer, D. G. 1995, 
\mnras, 272, 41

\bibitem[Stasi\'nska(1978)]{sta78} 
Stasi\'nska, G. 1978, 
\aap, 66, 257

\bibitem[Torres-Peimbert \& Peimbert(2003)]{tor03} 
Torres-Peimbert, S., \& Peimbert, M. 2003, 
in IAU Symposium 209, Planetary Nebulae and Their Role in the Universe, 
ed. S. Kwok, M. Dopita, \& R. Sutherland (San Francisco: 
ASP), 363

\bibitem[Tsamis et al.(2003)]{tsa03} 
Tsamis, Y.G., Barlow, M.J., Liu, X.-W., Danziger, I.J., \& Storey, P.J. 2003, 
\mnras, 338, 687

\bibitem[Verner et al.(2000)]{ver00} 
Verner, E.M., Verner, D.A., Baldwin, J.A., Ferland, G.J., \& Martin, P.G. 2000, 
\apj, 543, 831

\bibitem[Wiese, Smith, \& Glennon(1966)]{wie66} 
Wiese, W.L., Smith, M.W., \& Glennon, B.M. 1966, 
Atomic Transition Probabilities. (NBS 4)(Washington: NBS)

\bibitem[Wiese, Fuhr, \& Deters(1996)]{wie96} 
Wiese, W.L., Fuhr, J.R., \& Deters, T.M. 1996, 
in Atomic Transition Probabilities of Carbon, Nitrogen, and
Oxygen: A Critical Data Compilation, Journal of Physical and
Chemical Data, Monograph No. 7

\bibitem[Zhang(1996)]{zha96} 
Zhang H. L., 1996, 
A\&AS, 119, 523

\bibitem[Zhang \& Pradhan(1997)]{zha97} 
Zhang H. L., \& Pradhan A. K., 1997, 
A\&AS, 126, 373

\end{thebibliography}
\end{document}